\documentclass[12pt,letter]{article}

\usepackage{graphicx, epsfig, color, amsmath, amssymb}
\def\eslt{\not\!\!{E_T}}
\def\esltTwo{\not\!\!\!{E_T}}
\def\to{\rightarrow}

\def\bi{\begin{itemize}}
\def\ei{\end{itemize}}
\def\te{\tilde e}

\def\tu{\tilde u}
\def\sps1ap{SPS1a$^\prime$}
\def\c1p{C1$^\prime$}

\def\tb{\tilde b}

\def\tst{\tilde t}
\def\ttau{\tilde \tau}

\def\tg{\tilde g}
\def\tnu{\tilde\nu}

\def\tw{\widetilde W}
\def\tz{\widetilde Z}
\def\alt{\stackrel{<}{\sim}}
\def\agt{\stackrel{>}{\sim}}
\def\be{\begin{equation}}  
\def\ee{\end{equation}}  
\def\bea{\begin{eqnarray}}  
\def\eea{\end{eqnarray}}  
\def\beas{\begin{eqnarray*}}  
\def\eeas{\end{eqnarray*}}

\newcommand\epjc[3]{{\it Eur.\ Phys.\ J. }{\bf C #1} (#2) #3}

\begin{document}
\begin{titlepage}
\begin{flushright}
UH-511-1285-17
\end{flushright}

\vspace{0.5cm}
\begin{center}
{\Large \bf Aspects of the same-sign diboson signature \\
from wino pair production with light higgsinos \\
at the high luminosity LHC {\vspace{-12 pt}}}\\ 
\vspace{1.2cm} \renewcommand{\thefootnote}{\fnsymbol{footnote}}
{\large Howard Baer$^1$\footnote[1]{Email: baer@ou.edu }, 
Vernon Barger$^2$\footnote[2]{Email: barger@pheno.wisc.edu },
James S. Gainer$^3$\footnote[3]{Email: jgainer@hawaii.edu },\\
Michael Savoy$^1$\footnote[4]{Email: savoy@ou.edu },
Dibyashree Sengupta$^1$\footnote[5]{Email: Dibyashree.Sengupta-1@ou.edu },
and Xerxes Tata$^3$\footnote[6]{Email: tata@phys.hawaii.edu }
}\\ 
\vspace{1.2cm} \renewcommand{\thefootnote}{\arabic{footnote}}
{\it {\vspace{-12 pt}}
$^1$Dept. of Physics and Astronomy,
University of Oklahoma, Norman, OK 73019, USA \\
}
{\it 
$^2$Dept. of Physics,
University of Wisconsin, Madison, WI 53706, USA \\
}
{\it 
$^3$Dept. of Physics and Astronomy,
University of Hawaii, Honolulu, HI 96822, USA \\
}

\end{center}

\vspace{0.5cm}
\begin{abstract}
Naturalness arguments applied to simple supersymmetric (SUSY) theories
require a set of light higgsinos with mass $\sim |\mu|$ not too far from
$m_h$.  These models have an inverted electroweakino spectrum
with $|\mu| \ll M_2$ which leads to a rather clean, hadronically quiet,
same-sign diboson (SSdB) signature at hadron colliders arising from
neutral-plus-charged wino pair production.  
We improve and expand our earlier studies of this signature 
for discovering SUSY in natural SUSY models
by (i)~including backgrounds which were not
previously considered and which turn out to be significant,
(ii)~devising more efficient cuts to successfully contend with
these larger backgrounds and determining the discovery reach and
exclusion ranges for winos with these cuts, emphasizing
projections for the updated integrated luminosity target for HL-LHC of
3~ab$^{-1}$, and (iii)~emphasizing the utility of this channel for
natural models without gaugino mass unification.  We display the
kinematic characteristics of the relatively jet-free same sign
dilepton+$\esltTwo$ events (from leptonic decays of both $W$s) and find
that these are only weakly sensitive to the parent wino mass. We also
examine the charge asymmetry in these events and show that its
measurement can be used to check the consistency of the wino origin of
the signal. Finally, we show that -- because the wino branching fractions
in natural SUSY are essentially independent of details of the underlying
model -- a determination of the rate for clean, same-sign dilepton events
yields a better than 10\% determination of the wino mass over the entire
mass range where experiments at the HL-LHC can discover the wino signal.

\noindent 
\vspace*{0.8cm}

\end{abstract}

\end{titlepage}

\section{Introduction}
\label{sec:intro}

The search for supersymmetry in Run 2 of LHC with $\sqrt{s}=13$ TeV and
$\sim 36$ fb$^{-1}$ of data has resulted in mass limits of $m_{\tg}\agt
2$ TeV~\cite{lhc_mgl} and $m_{\tst_1}\agt 0.9$ TeV~\cite{lhc_mt1}.
These rather severe mass limits have led to concern that simple SUSY
models may be entering the regime of {\it unnaturalness}; if true, such
considerations could undermine the entire {\it raison d'etre} for weak
scale supersymmetry~\cite{craig}. It should, however, be stressed that
conclusions from naturalness regarding upper bounds on sparticle masses
\cite{bg,papp} (limits on stop masses are the most widely discussed)
do not apply if the model parameters-- often assumed to be independent-- 
turn out to be correlated\cite{reduces,am,seige}.

Quantitative measures of naturalness
generally derive from calculations of the {\it fine-tuning} of
the weak scale, typically represented by the $Z$ boson mass,
which is related to other weak-scale SUSY parameters via
the MSSM scalar potential minimization condition,
\be
\frac{m_Z^2}{2}=\frac{m_{H_d}^2+\Sigma_d^d-(m_{H_u}^2+\Sigma_u^u)\tan^2\beta}{\tan^2\beta -1}-
\mu^2\sim -m_{H_u}^2-\mu^2-\Sigma_u^u(\tst_{1,2}) .
\label{eq:mzs}
\ee 
where $m_{H_{u,d}}^2$ are soft SUSY breaking Higgs mass parameters,
$\mu$ is the superpotential Higgs/ higgsino mass term, $\tan\beta\equiv
v_u/v_d$ is the ratio of Higgs field vacuum expectation values (vevs),
and the $\Sigma_{u}^{u}$ and $\Sigma_d^d$ terms include a variety of radiative
corrections (expressions for these can be found in the Appendix of
Ref.~\cite{rns}).  
Recently, several of us have suggested using electroweak naturalness 
as a conservative criterion~\cite{rns, ltr} to determine whether a 
SUSY model spectrum is unnatural.
The {\it electroweak} naturalness measure is defined as
\be \Delta_{EW}=max|{\rm each\ term\ on\ the\ RHS\ of\
Eq.~\ref{eq:mzs}}|/(m_Z^2/2).
\label{eq:Del-EW}
\ee 
Naturalness, then, is the requirement that $\Delta_{EW}$ is
relatively small. Conservatively, requiring $\Delta_{EW}< 30$ implies:
\begin{itemize}
\item $|\mu | \sim 100-300$ GeV (the closer to $m_Z$ the better);
\item $m_{H_u}^2$ is radiatively driven from large high scale values to
small negative values ($\sim -(100-300)^2$ GeV$^2$) at the weak scale;
\item the magnitude of $\Sigma_u^u$ is also bounded by about
  (300~GeV)$^2$. This is possible even if stop masses -- though bounded
  above -- are in the multi-TeV range, and gluinos are as heavy as 5-6~TeV
  \cite{helhc} (depending on the details of the model).\footnote{The
  limit on the gluino mass arises because radiative corrections from
  gluino loops raise the stop mass, and as a result $\Sigma_u^u(\tst)$
  becomes too large \cite{sundrum}.}
\end{itemize}
These conditions are met in a class of ``Radiatively-driven Natural SUSY
models'' (RNS)~\cite{rns}.  In these SUSY models with low $\Delta_{EW}$,
the largest of the radiative corrections typically come from the
top-squark sector contributions to $\Sigma_u^u$ and are minimized for
highly mixed TeV scale top squarks, a condition which also lifts the
Higgs mass, $m_h$, into the vicinity of its measured value $\sim 125$
GeV~\cite{rns, ltr}.  We emphasize, however, that as Eq.~(\ref{eq:mzs})
holds in general in the MSSM, the argument that naturalness in the MSSM
leads to small $|\mu|$, and concomitantly light
higgsinos,\footnote{Here, we
are implicitly assuming that the superpotential parameter, $\mu$, is the
dominant source of the higgsino mass. A soft SUSY-breaking contribution
to the higgsino mass is possible if there are no additional gauge
singlets that couple to higgsinos ~\cite{ross}. In extended frameworks
with additional TeV scale fields it is theoretically possible to
decouple the higgsino mass from the Higgs boson mass parameter that
enters into Eq.~(\ref{eq:mzs})~\cite{other}. } applies whether or not
one uses Eq.~(\ref{eq:Del-EW}) to define fine-tuning.

We advocate using $\Delta_{EW}$ for discussions of naturalness.  It
yields a conservative measure of fine-tuning because it allows for the
possibility that model parameters, frequently regarded as independent,
might turn out to be correlated once the SUSY breaking mechanism is
understood. Ignoring this may lead to an over-estimate of the UV
sensitivity of $m_Z^2$ and cause us to prematurely discard perfectly
viable models. We also mention that the commonly used Barbieri-Giudice
measure \cite{bg, eenz} of fine-tuning reduces to $\Delta_{EW}$ once
appropriate correlations between model parameters are properly
implemented \cite{reduces,am}. That the use of $\Delta_{EW}$ to assess
naturalness is indeed conservative is brought home by explicit
examples~\cite{am} where the evaluation of $\Delta_{BG}$ with
parameter correlations ignored yields $\Delta_{BG} > 300\,\Delta_{EW}$.

While naturalness favors a small superpotential $\mu$ parameter, LHC
results seem to favor rather heavy gauginos, at least in models with
gaugino mass unification (where gaugino masses are related by
$M_1=M_2=M_3\equiv m_{1/2}$ at the energy scale $Q=m_{GUT}\simeq 2\times
10^{16}$ GeV).  In such models, renormalization group evolution of
gaugino masses typically leads to weak scale gaugino masses in the ratio
$M_1:M_2:M_3\sim 1:2:7$.  LHC limits on the gluino mass suggest
$M_3(weak)\agt 2$ TeV, which then implies that the wino mass, $M_2$,
$\agt 600$ GeV, and $M_1\agt 300$ GeV. We should, however, keep in mind
that gaugino mass unification is {\em not} a prerequisite for
naturalness~\cite{Baer:2015tva}, and also that direct limits from electroweak
gaugino searches at the LHC should be regarded as independent of those
from gluino searches.  Indeed searches for wino pair production
\cite{winosearch} in simplified models where the charged wino decays via
$\tw^\pm \to W^\pm$+ the lightest supersymmetric particle (LSP), and the
neutral wino decays via $\tw^0 \to Z$+LSP lead to  lower
bounds $\sim 500$~GeV for an LSP mass of about 200~GeV. Interestingly,
the strongest bound arises from the dilepton-plus-jet channel rather
than the clean but rate-suppressed trilepton channel. 
One might naively expect that as long as the higgsinos are 
essentially invisible these bounds will continue to apply.  
However, these bounds weaken
considerably in natural SUSY models once the expected branching
fractions (see below) for wino decays to light higgsinos are
incorporated, and there is essentially no bound if higgsinos are heavier
than about 150~GeV but still significantly lighter than the
winos.\footnote{While this is strictly speaking true only for the
analysis using chargino-neutralino production alone, in natural SUSY
chargino pair production also makes a (subdominant) contribution to the
$WZ$ channel. The upper limits on winos of natural SUSY will nonetheless
be significantly reduced from those in Ref.\cite{winosearch}.}

The inversion of the gaugino-higgsino mass pattern expected in natural
supersymmetry has important
implications not only for SUSY collider searches but also for dark
matter expectations. Since the lightest SUSY particle is
expected to be a higgsino-like neutralino, it is thermally
underproduced as dark matter.  Naturalness in the QCD sector
seems to require introduction of an axion~\cite{pqww} which may be
expected to constitute the remainder of the dark matter~\cite{bbc}.
While the axion and its cousins are well-motivated, we recognize that 
there are many other possibilities that could lead to the observed dark
matter, including out of equilibrium decays of heavy particles into the
neutralino LSP.

Though $M_3$ is phenomenologically constrained to be $\agt 2$~TeV,
without prejudices from gaugino mass unification the electroweak gaugino
mass parameters are relatively unconstrained.  If, motivated by
naturalness considerations, we assume $|\mu|$ is not hierarchically
larger than $M_Z$, then it is reasonable to explore LHC prospects
for SUSY scenarios with,
\be |\mu |< M_1, M_2 < M_3,
\label{eq:gauginos-2}
\ee 
where the heavier (wino-like) charginos and neutralinos decay to the
light higgsinos via $\tw_2^\pm \to \tz_{1,2} + W^\pm$, $\tw_2^\pm \to
\tw_1^\pm + Z,h$ and $\tz_4 \to \tz_{1,2}+Z,h$, $\tz_4 \to \tw_1^\pm
+W^\mp$.\footnote{In denoting the wino-like neutralino by $\tz_4$ we
  have implicitly assumed that the wino is heavier than the bino. This
  is not really a limitation to the analysis because the bino-like
  state couples rather weakly and so is phenomenologically relatively
  less important, as long as it is not the LSP.}

Although electroweak higgsino pair production processes $pp \to
\tz_i\tz_j, \tw_1\tz_i$ ($i,j=1,2$) have a large rate for higgsino
masses $\sim 150-300$~GeV, it is difficult to detect these above SM
backgrounds unless electroweak gauginos are fortuitously also much
lighter than required by naturalness \cite{Baer:2015tva}. 
However, for the generic situation with $|M_{1,2}| \gg |\mu|$, 
the higgsino spectra are very compressed, 
resulting in only relatively soft visible decay
products from $\tw_1,\tz_2$ decays and modest missing transverse energy.
One strategy for searching for light higgsinos at the LHC focuses on
higgsino pair production in association with a hard jet from initial
state QCD radiation which also serves as a trigger. Detailed studies
show that although it may be possible to obtain a ``signal statistical
significance of $5\sigma$'' above backgrounds after hard cuts, the $S/B$
ratio is just $\sim 1$\%. It appears to us unlikely that the systematic
errors on the QCD background could be reduced to this level~\cite{mono}.

The $S/B$ ratio can be greatly improved by requiring an additional low
invariant mass, same flavor, opposite sign soft dilepton pair from
$\tz_2\to\tz_1\ell^+\ell^-$ in these hard monojet events. It has been
shown that higgsinos up to 200-220~GeV would be detectable at the
$5\sigma$ level at LHC14, assuming 
an integrated luminosity of 1~ab$^{-1}$ \cite{llj}.\footnote{The
detection of pair production of light higgsinos at $e^+e^-$ colliders
with $\sqrt{s}>2m(higgsino)$ should also be
straightforward \cite{bmt,ilcgroup}, at least for higgsino mass gaps
larger than 10~GeV.} Note though that this search will not cover the
entire space of SUSY models with $\Delta_{EW}< 30$ even at the high
luminosity LHC.

There are several ways to search for superpartners in natural SUSY models.  
Old favorites like gluino pair production~\cite{mgluino} and
top-squark pair production~\cite{stop} remain as important search
channels, although now cascade decay events may contain occasional low
mass dilepton pairs arising from $\tz_2\to\tz_1\ell^+\ell^-$
decay~\cite{lhc,baris}.  We have already mentioned the search for soft
dileptons in events triggered by a hard monojet (or monophoton). Indeed,
the first limits from such a search have been presented by the CMS
collaboration in the $m_{\tz_2}$ vs.  $m_{\tz_2}-m_{\tz_1}$
plane~\cite{cms_llj}.

\begin{figure}[tbp]
\begin{center}
\includegraphics[width=10cm,clip]{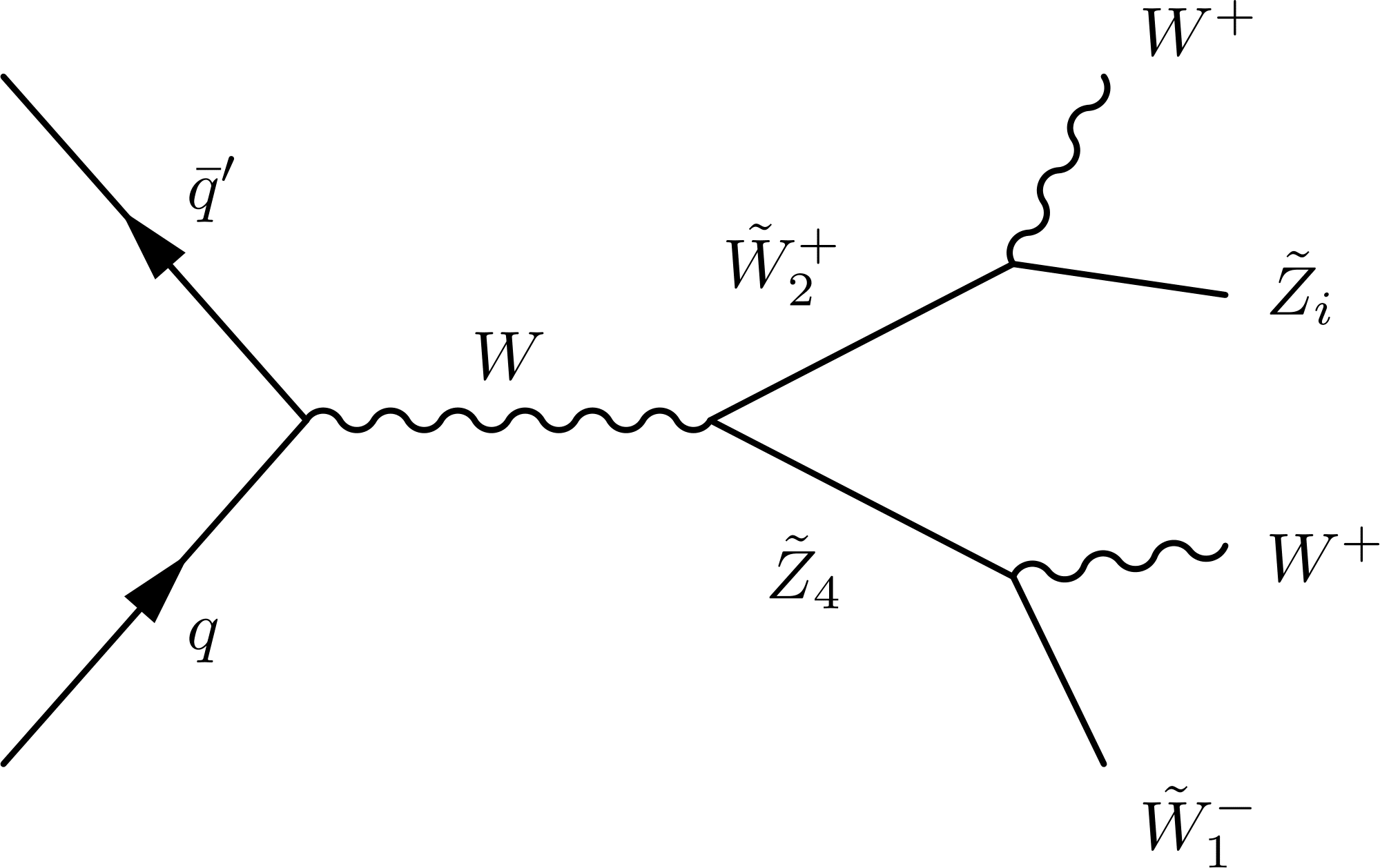}
\end{center}
\caption{A Feynman diagram for same-sign diboson production at LHC in
SUSY models with light higgsinos.
\label{fig:feyn}}
\end{figure}

Yet another distinctive signature for SUSY with light higgsinos (which
is the topic of this paper) arises from wino pair production~\cite{lhc,
lhcltr} via the Feynman diagram shown in Fig.~\ref{fig:feyn}:
$pp\to\tw_2^\pm\tz_4$ followed by $\tw_2^\pm\to W^\pm\tz_{1,2}$ and
$\tz_4\to W^\pm\tw_1^\mp$ decays.  Half of the time, the daughter $W$s
will have the same sign, leading to distinctive same sign di-boson (SSdB)
plus $\eslt$ events with no additional jet activity other than from QCD
radiation. The subsequent leptonic decays of the $W$s lead to {\it clean
same-sign dilepton} + $\eslt$ events for which the SM backgrounds are
very small. We stress that this class of same-sign dilepton events are
easily distinguished from those arising from gluino/squark pair
production~\cite{ssdl} because they are relatively free of accompanying
hard jet activity.

Some of us have examined this SSdB signature in previous
work~\cite{lhc, lhcltr}.  In these studies, the main SM backgrounds
considered were $t\bar{t}$, $WZ$, and $t\bar{t}W$ production (though
$t\bar{t}Z$ and inclusive $W^\pm W^\pm$ production from $qq \to
q'q'W^\pm W^\pm$ processes are also mentioned).  After a set of cuts
to help distinguish the natural SUSY SSdB signal from SM backgrounds,
it was found that the background dominantly arose from $t\bar{t}W$
production, and the LHC14 reach was obtained in the
two-extra-parameter non-universal Higgs (NUHM2)~\cite{nuhm2}
model\footnote{ Since the NUHM2 model allows the soft terms
  $m_{H_u}^2$ and $m_{H_d}^2$ to be traded for weak scale inputs $\mu$
  and $m_A$, it is easy to generate natural SUSY models by
  inputting low values of $|\mu| \sim 100-300$ GeV.}. It was emphasized
that in models with gaugino mass unification (such as the NUHM2
model), the SUSY reach  via the SSdB channel would (for integrated
luminosities larger than $\sim 100$~fb$^{-1}$) exceed the reach via
gluino pair production because the winos are only a third as light as
gluinos. This assumes that gluinos decay democratically to all
generations. In natural SUSY, where gluinos preferentially decay to
the third generation, it has been shown that $b$-tagging~\cite{btag}
could be used to further enhance the gluino reach~\cite{mgluino} in
the $\esltTwo$ channel. In Ref.~\cite{multi}, it was emphasized that for
natural SUSY models with gaugino mass unification, the
$pp\to\tz_1\tz_2 j$ reaction followed by $\tz_2\to\ell^+\ell^-\tz_1$
decay, combined with the SSdB channel, would cover the majority of
natural SUSY parameter space with $\Delta_{EW}<30$ at the high
luminosity LHC.  This conclusion no longer obtains in string-motivated
models such as natural generalized mirage mediation~\cite{ngmm} or the
minilandscape~\cite{mini} where the compressed spectrum of gauginos
may allow for both wino and gluino masses beyond HL-LHC reach even
while maintaining naturalness.

In the current paper, we revisit the SSdB signature from wino pair
production in SUSY models with light higgsinos, making a number of
important improvements.  First, we expand upon earlier calculations by
explicitly including several additional SM background processes:
(1) $WWjj$ production, (2) $t\bar{t}Z$ production, (3)
$t\bar{t}t\bar{t}$ production and (4) $WWW$ production.\footnote{ In
  addition, our current calculations adopt {\sc
    MadGraph}~\cite{madgraph} and {\sc Pythia}~\cite{pythia} for
  signal/background calculations and {\sc Delphes}~\cite{delphes} for
  our LHC detector simulation. While it is not obvious that {\sc
    Delphes}/{\sc PYTHIA} is an improvement over our previous use of
  the {\sc Isajet} detector simulation, the relative consistency of
  our new results with our previous results (when direct comparisons
  can be made) does provide a check on possible systematic errors.}
Second, we focus on the updated integrated luminosity target for the
HL-LHC, namely 3000 fb$^{-1}$ = 3 ab$^{-1}$.  Third, we emphasize that
the SSdB signature from wino pair production offers an 
{\it independent discovery channel} for natural SUSY models, whether
gaugino masses are unified or not.  For instance, in anomaly-mediated
SUSY breaking (AMSB) models, the gaugino masses are expected to occur
in the weak scale ratio of $M_1:M_2:M_3\sim 3.3:1:-7$. For natural
AMSB with $|\mu|\ll M_2$, it could be that gluino masses are well
above LHC reach while wino masses are quite light: $M_2\agt 300$
GeV. In such a case, the SSdB signature might be a robust discovery
channel even if gluinos are too heavy to be detected. Since we do  
not assume gaugino mass unification, we present
results in terms of the physical wino mass rather than 
{\it  e.g.} in terms of $m_{1/2}$.

In addition to presenting projections for the $5\sigma$ reaches for the
discovery of winos in this channel for various values of the wino mass
$m_{\tw_2}$ and the values of $m_{\tw_2}$ that can be expected to be excluded
at 95\% confidence level, we also analyze the prospects for wino mass
measurement.  We point out that using rate information, we can measure
the wino mass at better than the 10\% level over its entire discovery
range. We show that if there is an excess in the clean SS dilepton
sample, a determination of the charge asymmetry would provide an
important consistency check.
We also examine various kinematic distributions that may reveal
characteristic features of the SSdB events. 
We find that although these distributions in themselves are not 
strongly sensitive to the wino mass, they may
still be useful in a multivariate approach for extracting $M_2$. 
 
 We discuss our calculation of wino pair production, along with the
expected wino decay patterns in natural SUSY and describe our simulation
of signal and background processes in Sec.~\ref{sec:csec}. The analysis
cuts that we suggest for optimizing the SSdB signal at the HL-LHC are
described in Sec.~\ref{sec:cuts}. In Sec.~\ref{sec:reach} we show our 
projections of the discovery and exclusion reach for winos in the SSdB
channel, while various characteristics of signal events are discussed in
Sec.~\ref{sec:char}. In Sec.~\ref{sec:mass}, we examine the precision
with which the wino mass may be extracted from the SSdB signal rate.  Our
conclusions are presented in Sec.~\ref{sec:conclude}.

\section{Evaluation of signal and background cross sections}
\label{sec:csec}

\subsection{Signal production cross sections}
\label{ssec:signal}

\begin{figure}[tbp]
\begin{center}
\includegraphics[width=14cm,clip]{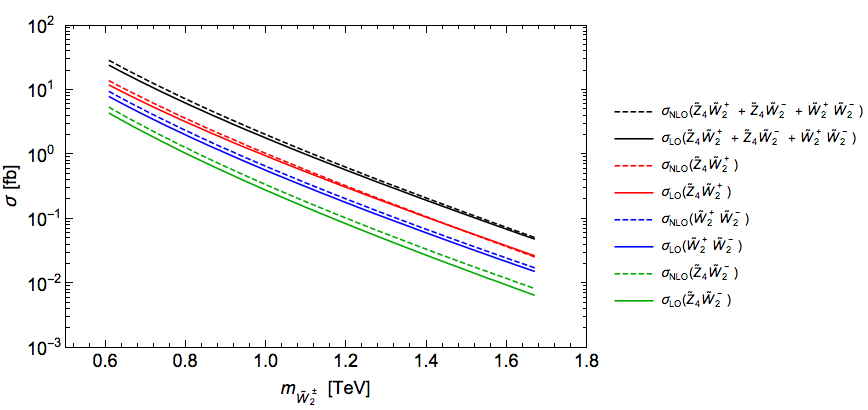}
\end{center}
\caption{Leading order (solid) and next-to-leading order (dashed)
  cross sections for various wino pair production processes at the LHC
  with $\sqrt{s}=14$ TeV.
\label{fig:sigma}}
\end{figure}

Since the SSdB signature from pair production of winos is the subject of
this study, we begin by showing in Fig. \ref{fig:sigma} the leading
order (LO) and next-to-leading order (NLO) production cross sections for
various wino pair production processes-- as solid and dashed curves
respectively. 
These cross
sections are calculated for the $\sqrt{s}=14$ TeV LHC using the {\sc
Prospino} computer code~\cite{prospino} and are plotted with respect to
the charged wino mass, $m_{\tw_2}$. Since we will also be interested in
examining the lepton charge asymmetry, we also show separately the cross
sections for $pp\to\tw_2^+\tz_4$ (red curves) and for
$pp\to\tw_2^-\tz_4$ (green curves).

Note that the $\tw_2^+\tz_4$ cross section typically exceeds the cross
section for $\tw_2^-\tz_4$ by a factor $\sim 3-4$. This charge
asymmetry in production cross section arises from the preponderance of
valence $u$ quarks in the proton versus valence $d$ quarks and
increases with $m_{\tw_2}$ due to the growing importance of valence
quark over sea quark annihilation as the sampled parton fractional
momentum, $x_F$, increases. This results in a preponderance of $++$ over $--$
dilepton events as we shall see below.

The charged wino pair production cross section $pp\to\tw_2^+\tw_2^-$ (blue curves)
lies in between the $\tw_2^+\tz_4$ and $\tw_2^-\tz_4$ curves. 
The black curves denote the cross sections for the 
summed wino pair production channels, which vary from the tens of fb 
level for $m_{\tw_2}\sim 600$ GeV to $\sim 10^{-2}$ fb
for $m_{\tw_2}\sim 1.6$ TeV.
\subsection{Wino branching fractions}
\label{ssec:signal-branching}

The $\tw_2$ and $\tz_4$ branching fractions are calculated using {\sc
  Isajet} 7.85~\cite{isajet} and have been shown in
Ref.~\cite{lhc, lhcltr}.  We remind the reader that for natural SUSY
with light higgsinos, the branching ratios for
$\tw_2^+\to\tz_{1,2}W^+,\ \tw_1^+Z$ and $\tw_1^+h$  decays
each rapidly asymptote to $\sim 25$\% for heavy winos with only
small branching fractions to the bino-like $\tz_3$. Likewise,
the branching fractions for $\tz_4\to \tw_1^+W^-$,
$\tw_1^- W^+$, $\tz_{1,2} Z$ and $\tz_{1,2} h$ are also each $\sim 25\%$
for $|\mu| \ll |M_2|$.

These simple decay patterns can be analytically understood in the limit
that the $\tw_1$ and $\tz_{1,2}$ are mostly higgsino-like, and $\tw_2$
and one of $\tz_3$ or $\tz_4$ is mostly a wino (with the other
neutralino being dominantly a bino). As already mentioned, the bino-like
neutralino couples to the wino only via its small higgsino component, so
decays to it are dynamically suppressed even if they are kinematically
allowed. In natural SUSY, we are interested in the case $\mu^2 \ll
M_2^2$, and medium to large $\tan\beta$ values, typically with $\tan\beta >
|M_2/\mu|$. In this case, it is straightforward to check that the
chargino mixing angle $\gamma_L \sim -\gamma_R \frac{\mu}{M_2}$ (we use
the notation of Ref.~\cite{wss}) so that $\gamma_L$ can be ignored
compared to $\gamma_R$. The small gaugino components of the
higgsino-like states and the higgsino components of the wino-like states
can be evaluated to lowest order in the gaugino-higgsino mixing angles,
and the relevant couplings and partial widths for the various decays
obtained from the expressions in Appendix B of Ref.~\cite{wss}.  We then
find
\begin{equation}
\Gamma(\tw_2\to \tz_1W)\simeq  \Gamma(\tw_2 \to \tz_2 W)\simeq 
\Gamma(\tw_2 \to \tw_1Z) \simeq  \Gamma(\tw_2 \to \tw_1 h) \simeq 
\frac{g^2}{64\pi}m_{\tw_2}, \end{equation}

\begin{equation}
\Gamma(\tz_4\to \tw_1^-W^+)\simeq  \Gamma(\tz_4 \to \tw_1^+ W^-)\simeq 
\Gamma(\tz_4 \to \tz_{1,2}Z) \simeq  \Gamma(\tz_4 \to \tz_{1,2} h) \simeq 
\frac{g^2}{64\pi}m_{\tz_4},
\label{eq:widths}
\end{equation}
where, to illustrate our point, we have retained only the largest mass
terms in the expressions for the partial widths. This is a good
approximation when higgsinos are much lighter than the winos. In our
numerical calculation, we retain the full expressions, of course.  In
the last of these equations we have assumed that $\tz_4$ is the
wino-like state. Also, the neutral wino decay widths to $Z$ or $h$ are
the summed widths to both higgsino-like states.\footnote{The reader may
wonder why the decay rates to Higgs bosons which go via the {\em
unsuppressed} wino-higgsino-Higgs boson coupling are comparable to the
decay rates to vector bosons which can only occur via small mixing
angles. The reason is that this suppression is compensated by the
enhancement of the amplitude for decays to longitudinal $W$ or $Z$
bosons by a factor $m_{\tw_2,\tz_4}/M_{W,Z}$, an example of the
Goldstone boson equivalence theorem. } If other decay modes of the wino
({\it e.g.}, to the bino, to sfermions, or to the heavy Higgs bosons) are
kinematically or dynamically suppressed, we obtain the approximately equal
branching fractions of 25\% mentioned above. We have checked by a
numerical scan that when $|\mu| = 150-300$~GeV, as favored by naturalness, 
the branching ratios for these modes are {\em well within} 
the 0.23-0.27 range if the wino is heavier than 500~GeV and the
bino is not quasi-degenerate with the wino.

Combining decay channels, we find that typically $\sim 1/8$ of
$\tw_2^\pm\tz_4$ production events lead to final states with same-sign
dibosons $W^+W^+$ or $W^-W^-$. To identify SSdB events, we require
leptonic decays of the final state $W$s to $e$ or $\mu$ which reduces 
our overall branching fraction to $\sim 6 \times 10^{-3}$.
Thus, although the wino pair production cross sections may be as large as
10~fb, the combined signal channel branching fractions lead to
relatively small signal rates. Therefore, the SSdB signal channel
really becomes the signal of choice only for the very high integrated
luminosities projected to be accumulated at the high-luminosity LHC.

\subsection{Signal benchmark model line}
\label{ssec:model-line}

To make specific predictions for the expected SSdB signal rate, we
will adopt a natural SUSY model line using the
two-extra-parameter non-universal Higgs model NUHM2~\cite{nuhm2}.
This model allows for direct input of a low $\mu$ parameter as
required by naturalness.  The model line we adopt is 
adapted from Ref.~\cite{lhc} and has $m_0=5$ TeV,
$A_0=-8$ TeV, $\tan\beta =10$, $m_A=1.5$ TeV, and $\mu =150$ GeV.  We
will allow the unified gaugino mass parameter $m_{1/2}$ to vary from
700 to 1375 GeV which corresponds to $m_{\tg}\sim 1.8-3.2$ TeV or
$m_{\tw_2}\sim 610-1200$ GeV. The value of $m_h$ is $\sim 125$ GeV along the
entire model line, while $\Delta_{EW}$ is $\sim 10-30$,  corresponding to
10\% - 3\% EW fine-tuning.  Although the NUHM2 model assumes a
unification of gaugino mass parameters, this is unimportant for the
analysis of the wino signal that we are focussing upon, in the sense
that essentially identical results would be obtained in any model with
the same value of the wino mass $M_2$.  While there may be some
sensitivity to the bino mass parameter, we remind the reader that the
bino-like state couples to the wino-vector boson system only via its
small higgsino components, so any decays into this state typically
have small branching fractions.

%
\begin{table}[htp]\centering
\begin{tabular}{lc}
\hline
parameter & {\tt Point B} \\
\hline
$m_0$      & 5000  \\
$m_{1/2}$       & 800  \\
$A_0$    & -8000  \\
$\tan\beta$      & 10  \\
$\mu$          & 150   \\
$m_A$          & 1500  \\
\hline
$m_{\tg}$   & 2007.4  \\
$m_{\tu_L}$ & 5170.2  \\
$m_{\tu_R}$ & 5318.4  \\
$m_{\te_R}$ & 4815.2  \\
$m_{\tst_1}$& 1470.3  \\
$m_{\tst_2}$& 3651.2  \\
$m_{\tb_1}$ & 3682.7  \\
$m_{\tb_2}$ & 5051.2  \\
$m_{\ttau_1}$ & 4740.2  \\
$m_{\ttau_2}$ & 5075.6  \\
$m_{\tnu_{\tau}}$ & 5082.8  \\
$m_{\tw_2}$ & 692.2  \\
$m_{\tw_1}$ & 155.2  \\
$m_{\tz_4}$ & 703.1  \\
$m_{\tz_3}$ & 363.1  \\
$m_{\tz_2}$ & 158.2  \\
$m_{\tz_1}$ & 142.4  \\
$m_h$       & 124.4  \\
\hline
$\Omega_{\tz_1}^{std}h^2$ & 0.008  \\
$BF(b\to s\gamma)\times 10^4$ & $3.1$  \\
$BF(B_s\to \mu^+\mu^-)\times 10^9$ & $3.8$ \\
$\sigma^{SI}(\tz_1, p)$ (pb) & $4.1\times 10^{-9}$  \\
$\sigma^{SD}(\tz_1 p)$ (pb) & $1.5\times 10^{-4}$  \\
$\langle\sigma v\rangle |_{v\to 0}$  (cm$^3$/sec)  & $2.9\times 10^{-25}$ \\
$\Delta_{\rm EW}$ & 9.3 \\
\hline
\end{tabular}
\caption{Input parameters and masses in~GeV units for an NUHM2 model
  SUSY benchmark point labeled {\tt Point B} with $m_t=173.2$ GeV and
  $m_{1/2}=800$ GeV.  }
\label{tab:bm}
\end{table}

In Table \ref{tab:bm}, we show a listing of various sparticle masses and
observables associated with our model line for the benchmark model with
$m_{1/2}=800$ GeV, labeled as {\tt Point B}.\footnote{We refer to this as
{\tt Point B} because we consider three signal benchmark points, labeled
A, B, and C, in order of increasing wino mass.}  Within the NUHM2
framework, the model point with the 692~GeV wino state $\tw_2$ has
$m_{\tg} \approx 2000$ GeV and so is just beyond the current gluino mass
limit (from $13$ TeV LHC running with $\sim 35$ fb$^{-1}$).  Though the
details of most of the SUSY spectrum are unimportant for our present
purposes, we note that our sample case (indeed the entire model line)
has very heavy first/second generation sfermions, with stops and gluinos
in between these and the EW gauginos, while higgsinos are very
light. This qualitative pattern is a generic feature of natural SUSY
models. We emphasize that while our benchmark model line is in a model
with gauge coupling unification, this will have very little (if any)
effect on any conclusions we draw about the prospects for discovery,
exclusion, or mass measurement of the parent wino. In other words, for
the purposes of analysis of the wino signal alone, we can disregard the
LHC gluino limit and model cases with lighter winos that may arise in 
natural models without gaugino mass unification
using $m_{1/2}$ as a surrogate for the wino mass, $M_2$.

\subsection{SM background cross sections}
\label{ssec:BG}

In order to assess prospects for observability of the signal, we must
have a good understanding of various SM backgrounds that could 
also lead to the clean same sign dilepton plus $\eslt$ signature. 
We have considered backgrounds from $t\bar{t}$, $WZ$, $t\bar{t}W$, $t\bar{t}Z$,
$t\bar{t}t\bar{t}$, $WWW$, and $W^\pm W^\pm jj$ production processes in
the SM. Top pair production yields (non-instrumental) backgrounds only
if a secondary lepton from top decay is accidently isolated. We use LO
event generation from {\sc MadGraph} in our simulation of both signals
and backgrounds, but rescale the LO total cross sections to be in accordance
with NLO values found in the literature.

Specifically, we use 953.6 pb as the total NLO cross section for
$t\bar{t}$, following Ref.~\cite{Czakon:2013goa}.
Ref.~\cite{Bevilacqua:2012em} gives us a K factor of $1.27$ for
four-top production.  We use $1.88$ as the K factor for associated
$WZ$ production following Ref.~\cite{Campbell:2011bn} and $1.24$ for
the K factor for $t\bar{t}W$ production following
Ref.~\cite{Campbell:2012dh}\footnote{ While in
  Ref.~\cite{Campbell:2011bn}, K factors differ slightly for $W^+Z$
  and $W^-Z$, and in Ref.~\cite{Campbell:2012dh} the K factors differ
  slightly for $t\bar{t}W^+$ and $t\bar{t}W^-$, these are very close
  ($1.86$ and $1.92$ respectively for $W^+Z$ and $W^-Z$ and $1.22$ and
  $1.27$ for $t\bar{t}W^+$ and $t\bar{t}W^-$ respectively), especially
  when compared with likely theory errors, so we use $1.88$ ($1.24$)
  as the K factor for both $WZ$ ($t\bar{t}W$) processes.}.  We obtain
the K factor $1.39$ for $t\bar{t}Z$ from Ref.~\cite{Kardos:2011na};
Ref.~\cite{Melia:2010bm} gives us a K factor of $1.04$ for
$WWjj$\footnote{ This is the value in Ref.~\cite{Melia:2010bm} for the
  two-jet inclusive cross section with factorization and
  renormalization scales set to $150$ GeV.  If we were to further
  restrict to one-jet and zero-jet bins (see our analysis cuts,
  below), the K factor would move closer to $1$; we have chosen the
  larger K factor to be conservative.}.  Finally, for the $WWW$
process we use the cross sections in Ref.~\cite{Yong-Bai:2016sal}.  In
our analyses we use a common K factor of $2.45$ for both $WWW$
processes, which is not appreciably different than the $W^+W^+W^-$ K
factor of $2.38$ or the $W^+W^-W^-$ K factor of $2.59$.  We note that
these are K factors for inclusive $WWW$ production; if one imposes a
jet veto the K factor is significantly reduced (to $1.29$ for the
combined $WWW$ K factor).  While we do impose a jet multiplicity cut
of $n_{jet} \le 1$, we choose to be conservative and use the larger
value for the K factor in our calculation of the background.

These K factors and NLO cross sections for the underlying fundamental SM
processes 
are shown  in columns 2 and 3 of Table~\ref{tab:bg}, together with the
corresponding information for the signal benchmark {\tt Point B}. These
are, of course, the raw production cross sections for the various final
states; various branching fractions and detection efficiencies have
to be folded in to obtain the signal and background cross sections. We
see that even the various $2\to 3$ and $2\to 4$ SM processes have
potentially larger rates than the signal, so we may anticipate that
we will
require relatively stringent selection cuts to make the signal observable.

\subsection{Event simulation}
\label{ssec:simulation}

To simulate SSdB signal events, we first generate the SUSY spectrum as
a Les Houches Accord (LHA) file using {\sc Isajet} 7.85~\cite{isajet}.
We then feed the LHA information to {\sc MadGraph/ MadEvent}
2.3.3~\cite{madgraph} which is interfaced with {\sc Pythia}
6.4~\cite{pythia} for parton showering and hadronization.  The
generated events are passed to {\sc Delphes} 3.3.0~\cite{delphes} 
for fast detector simulation, 
where we utilize the default ``CMS'' parameter card for
version 3.3.0 with the modifications listed below.

\begin{enumerate}

\item We require jets to have transverse energy $E_T(jet) > 50$ GeV
and pseudorapidity $|\eta(jet)| < 3.0$. 

\item The electromagnetic calorimeter (ECAL) energy resolution is set to
  $3\%/\sqrt{E} \oplus 0.5\%$, while the hadronic calorimeter (HCAL) energy
  resolution is taken to be $80\%/\sqrt{E} \oplus 3\%$ for $|\eta| < 2.6$ and
  $100\%/\sqrt{E} \oplus 5\%$ for $|\eta| > 2.6$, where $\oplus$ denotes
  combination in quadrature. 

\item The jet energy scale correction is turned off.

\item The anti-$k_T$ jet algorithm~\cite{Cacciari:2008gp}
is utilized, but using $R = 0.4$ rather than the default $R = 0.5$.  
(Jet finding in Delphes is implemented via 
{\sc FastJet}~\cite{Cacciari:2011ma}.)  
 One motivation for choosing $R = 0.4$ in the jet algorithm is to facilitate
comparison with CMS $b$-tagging efficiencies~\cite{CMS:2016kkf}.

\item We performed jet flavor association using our own module which
implements the ``ghost hadron'' procedure~\cite{Cacciari:2007fd} which
allows the assignment of decayed hadrons to jets in an unambiguous manner.
We use this module to aid in $b$-tagging, specifically in determining
whether jets contain $B$ hadrons.  
When a jet contains a $B$ hadron in which the $b$ quark will decay at the next
step of the decay, then if this $B$ hadron lies within $|\eta| < 3.0$ 
and $E_T > 15$ GeV, we identify this $b$-jet as a ``truth $b$-jet''.  
We $b$-tag truth $b$-jets with $|\eta| < 1.5$ with an efficiency of $60\%$.
We also $b$-tag jets which are not truth $b$-jets 
with $|\eta| < 1.5$ with an efficiency of $1/X$
where $X = 150$ for $E_T < 100$ GeV, 
$X = 50$ for $E_T > 250$ GeV
and $X$ is found from a linear interpolation for $100$ GeV $ < E_T < 250$
GeV\footnote{The parameters for this $b$-tagging procedure 
are based on ATLAS studies of $b$-tagging
efficiencies and rejection factors in $t\bar{t}H$ and $WH$ production
processes~\cite{ATLASb}.}.  We have checked~\cite{mgluino} that our $b$-jet
tagging algorithm yields good agreement with the $b$-tagging
efficiencies and mistag rates in Ref.~\cite{CMS:2016kkf}; 
specifically it gives results intermediate between the CMS ``medium''
and ``tight'' $b$-tagging algorithms.

\item ``Tau tagging'', {\it i.e.}, identifying objects as taus, is not
used.

\item The lepton isolation modules were modified to allow us
to adopt the isolation criterion that the sum of $E_T$ of physics
objects in a cone with $\Delta R < 0.2$ about the lepton
direction is less than min$(5$ GeV, $0.15 E_T(\ell))$,
where $E_T(\ell)$ is the transverse energy of the lepton.
({\sc Delphes} 3.3.0 did not allow the minimum of these
two thresholds to be used rather than using either
a fixed value of $E_T$ or a fraction of the lepton $E_T$.)

\end{enumerate}

\section{Analysis cuts to enhance SUSY SSdB signal}
\label{sec:cuts}

\subsection{Initial selection cuts ({\bf C1})}
\label{ssec:C1}

We begin by imposing the selection cuts, listed below, that were
suggested in
Ref's.~\cite{lhcltr, lhc} to enhance same sign dilepton events
originating in wino production over those coming from SM processes.
\bi
\item Exactly two isolated same-sign leptons with $p_T(\ell_1)>20$ GeV
and $p_T(\ell_2)>10$ GeV.  ($\ell_1$ denotes the higher $p_T$ lepton, while
$\ell_2$ is the lower $p_T$ lepton.)
\item $n(b{\rm-jets})=0$ 
\item $\eslt >200$ GeV, and
\item $m_T^{min}>175$ GeV,
\ei
where $m_T^{min}=min [m_T(\ell_1,\eslt ,m_T(\ell_2,\eslt)]$.
We denote these initial cuts as cut set {\bf C1}.

The cross sections after these cuts-- after folding in various
branching fractions and detection efficiencies-- for the {\tt Point B} signal
benchmark point and from various SM processes (in ab) are listed in
column 4 of Table \ref{tab:bg}.  The combined same-sign dilepton cut,
large $\eslt$ cut, and $b$-jet veto serve to severely reduce the
$t\bar{t}$ background. Indeed, after these cuts, the analysis of
Ref.~\cite{lhc, lhcltr} found the dominant background to come from
$t\bar{t}$ and $WZ$ production.  Any $t\bar{t}$ background events which
survive these cuts will likely have one lepton arising from real
$W\to\ell\nu$ decay with the other lepton arising from a semi-leptonic
$b$ decay, which will hence be soft. In such a case, at least to the extent that
the $\eslt$ dominantly arises from the leptonic decay of a single $W$,
the transverse mass, $m_T(\ell ,\nu_\ell )$, is mostly bounded by $m_W$
(up to small contamination from off-shell $W$s, $\eslt$ smearing, and any
additional $\eslt$ from leptonic decays of the $B$-hadron).  Thus, the
further requirement of $m_T^{min}\gg m_W$ should serve to greatly reduce the
$t\bar{t}$ and also $WZ$ backgrounds. Here, in accord with
Refs.~\cite{lhc, lhcltr}, we require $m_T^{min}>175$ GeV; after imposing
this cut we are indeed left with no $t\bar{t}$ or $WZ$ backgrounds in
our samples.  Among the largest backgrounds is $t\bar{t}W$ production,
which we find to be a factor of two larger than in
Ref.~\cite{lhc}. Unlike the earlier studies, we also find sizable
contributions from $t\bar{t}Z$ production as well as from $WWW$
production
and $W^\pm W^\pm jj$ production.
Summing these sources, we find a total background cross section after
{\bf C1} cuts of 34 ab in contrast to just 6 ab after the same cuts in
Ref.~\cite{lhc}. The cross section for the signal at the benchmark
{\tt Point B} is 29 ab, or a little under $5\sigma$ statistical
significance for an integrated luminosity of 1~ab$^{-1}$, and over
$8.5\sigma$ significance with 3~ab$^{-1}$.

\subsection{Optimizing the reach of HL-LHC: selection cuts {\bf C2}}
\label{ssec:C2}

The cut set {\bf C1} was suggested in Ref.~\cite{lhc, lhcltr} to determine
the reach of LHC14 in the SSdB channel for 100-1000 fb$^{-1}$.
Since one of our goals is to project the maximum reach of the
HL-LHC for SUSY in the SSdB channel, we attempt to further optimize
our cuts.

\begin{figure}[tbp]
\begin{center}
\includegraphics[width=14cm,clip]{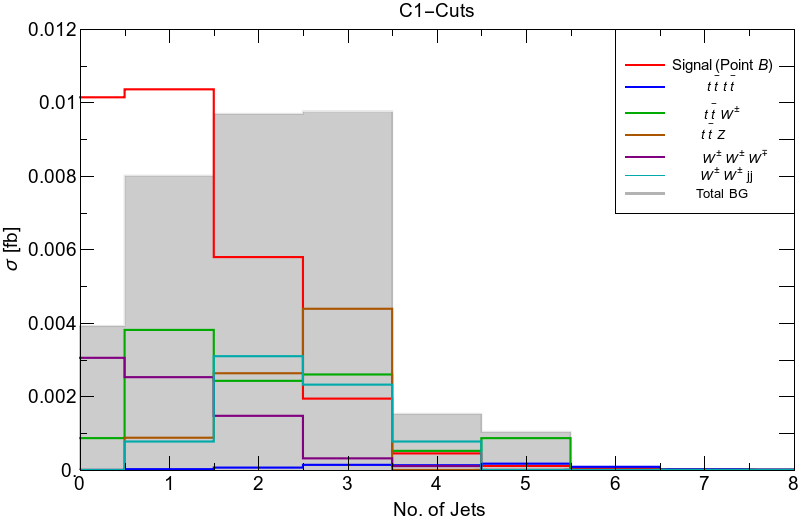}
\end{center}
\caption{Distribution of jet multiplicity, $n(j)$, for SSdB events
  from the {\tt Point B} signal benchmark point and various SM
  backgrounds after {\bf C1} cuts.
\label{fig:njets}}
\end{figure}

We begin by noting that the various background processes in
Table~\ref{tab:bg} with significant cross sections after {\bf C1} cuts
are all expected to contain additional hard jets, while jet activity in
the signal process arises only from initial state QCD radiation (and
very soft jets from decay of the heavier higgsinos). We thus anticipate
that jet multiplicity will be a useful discriminating
variable.\footnote{In this vein, the scalar sum of jet $E_T$ or the
ratio of this to the scalar sum of leptonic $E_T$ may prove to be even
more robust and equally discriminating variables. } With this motivation
we show the expected jet multiplicity, $n(j)$, from signal and
background events after the {\bf C1} cuts in Fig. \ref{fig:njets}.  From
the solid (red) signal histogram, we see that signal events indeed
mainly have $n(j)=0$ or 1.  In contrast, background events, the sum of
which is
shown by the shaded histogram, generally have $n(j)\ge 2$.  Thus, we
apply the additional cut, \bi
\item $n(j)\le 1$.
\ei

\begin{figure}[tbp]
\begin{center}
\includegraphics[width=15cm,clip]{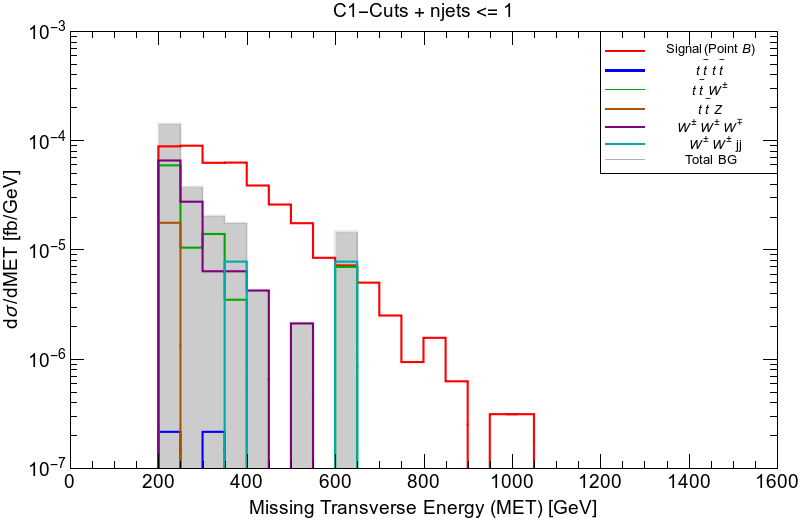}
\end{center}
\caption{Distribution of $\eslt$ for 
the  signal benchmark {\tt Point B} and various SM backgrounds in SSdB
production after {\bf C1} cuts plus the $n(j)\le 1$ cut.
\label{fig:met}}
\end{figure}

The cross sections 
after cut set {\bf C1} and $n(j)\le 1$ are listed in column 5 of
Table~\ref{tab:bg}. We see that the main background contributions 
now come from $t\bar{t}W$ and $WWW$ production processes. 
To further reduce these, we examined several other kinematic
distributions including $\eslt$, $m_T(\ell_1\ell_2,\eslt)$ (the
dilepton-plus-$\eslt$ cluster transverse mass)~\cite{mct}, $m_T^{min}$
and $m_{T2}$~\cite{mt2}.  The most useful of these turned out to be
the $\eslt$ distribution shown in Fig. \ref{fig:met}.
From this figure, we see that in the $\eslt=200-250$ GeV bin, the summed
background exceeds the  signal for {\tt Point B}, while in higher
$\eslt$ bins, signal clearly emerges above background. However, care
must be taken since our signal rate is already
rather small. We elect to make one final cut
\bi
\item $\eslt >250$ GeV,
  \ei
and label this set of cuts ({\bf C1} cuts plus $n(j) \le 1$, plus
$\eslt>250$ GeV) as the cut set {\bf C2}.

\begin{table}[htp]
\begin{center}
\begin{tabular}{|c|c|c|c|c|c|}
\hline process & $K-$factor & $\sigma$(NLO) (ab) & {\bf C1} & {\bf C1}
$+~n_{jet} \le 1$ & {\bf C2} \\ \hline \hline SUSY\ ({\tt Point B}) &
1.25 & $1.55\cdot 10^{4}$ & 28.8 & 20.5 & 16.1 \\ \hline $t\bar{t}$ &
1.72 & $9.5\cdot 10^{8}$ & 0 & 0 & 0 \\ $WZ$ & 1.88 & $5.2\cdot
10^{7}$ & 0 & 0 & 0 \\ $t\bar{t}W$ & 1.24 & $5.2\cdot 10^{5}$ & 11.1 &
4.7 & 1.7 \\ $t\bar{t}Z$ & 1.39 & $8.8\cdot 10^{5}$ & 7.9 & 0.9 & 0
\\ $t\bar{t}t\bar{t}$ & 1.27 & $1.1\cdot 10^{4}$ & 0.6 & 0.  & 0.
\\ $WWW$ & 2.45 & $3.2\cdot 10^{5}$ & 7.4 & 5.6 & 2.3 \\ $WWjj$ & 1.04
& $3.9\cdot 10^{5}$ & 7.0 & 0.8 & 0.8 \\ \hline total BG & -- &
$1.0065\cdot 10^{9}$ & 34.1 & 11.9 & 4.8 \\ \hline
\end{tabular}
\caption{Component background and signal cross sections in $ab$ before
  any cuts, after {\bf C1} cuts, after {\bf C1} cuts plus a jet veto, and
  after {\bf C2} at LHC14. Also shown is the $K$-factor that we use. 
\label{tab:bg}}
\end{center}
\end{table}

\begin{figure}[htbp]
\begin{center}
\includegraphics[width=8 cm,clip]{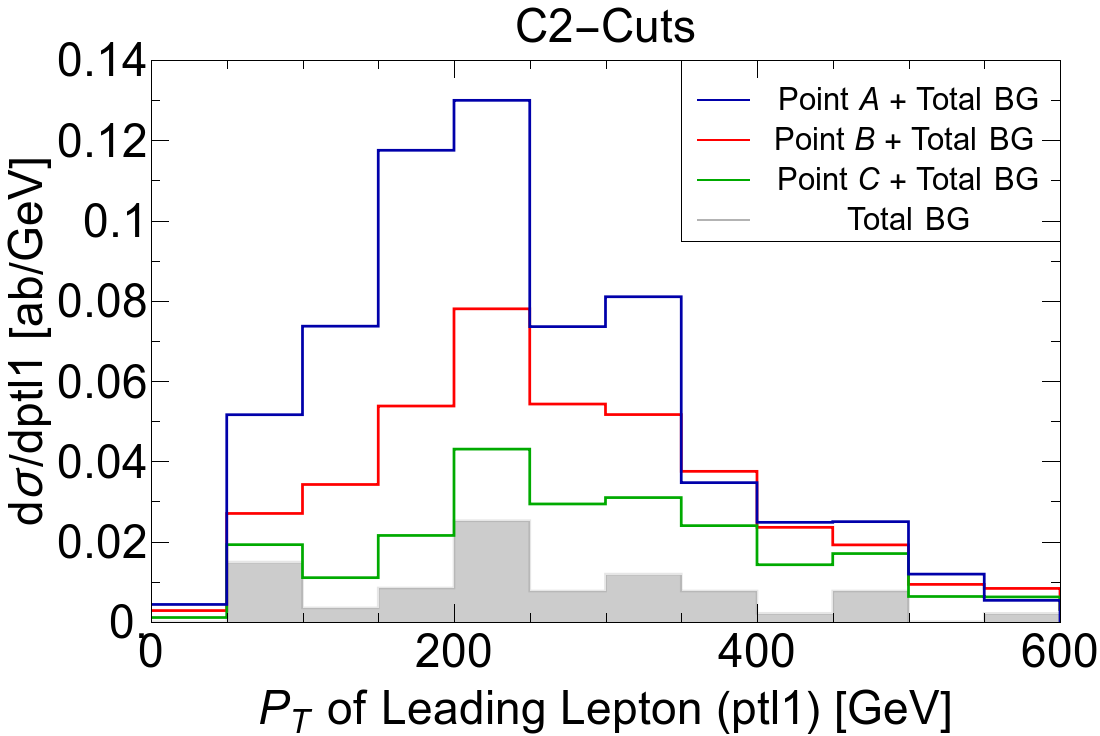}
\includegraphics[width=8 cm,clip]{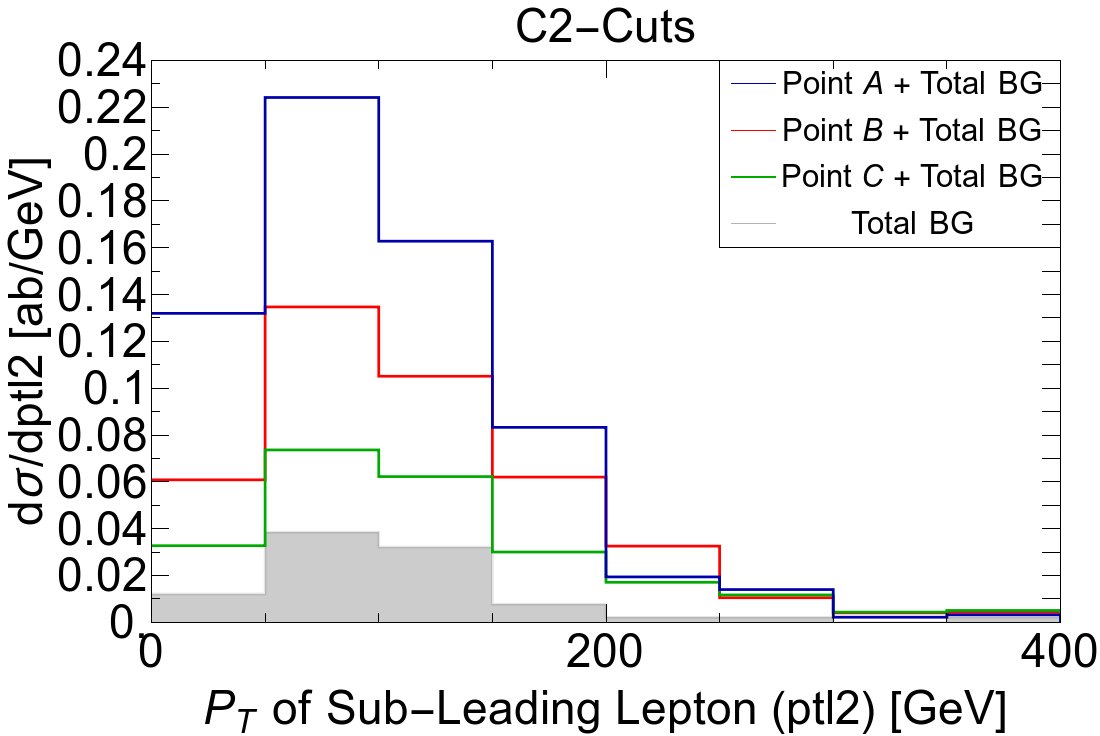}
\end{center}
\caption{Distribution of $p_T(\ell_1 )$ (left frame) and $p_T(\ell_2 )$
  (right frame ) for the {\tt Point A}, {\tt Point B}, and {\tt Point C}
  benchmarks, which are points along our NUHM2 model line with
  $m_{\tw_2} =$ 530, 692 and 886~GeV, respectively, together with the total SM
  background after {\bf C2} cuts.
\label{fig:ptl1}}
\end{figure}

We show the expected $p_T$ distributions of the leptons after the {\bf
C2} cuts in Fig.~\ref{fig:ptl1} for three signal benchmark points along
the model line, as well as for the summed SM background.  The points have
$m_{\tw_2}= 530$~GeV ({\tt Point A}), 692~GeV ({\tt Point B}, already
introduced above), and 886~GeV ({\tt Point C}).
We see that the distributions are qualitatively similar, and while the
$S/B$ ratio may be slightly improved by requiring harder cuts on the
leptons, this would only be at the cost of reducing an already
rate-limited signal. We choose, therefore, not to impose any further cuts.

The total background after these cuts is shown in the last column of
Table~\ref{tab:bg}. We see that almost half this background comes from
SM $WWW$ production. We remind the reader of our discussion in
Sec.~\ref{ssec:BG}, where we mentioned that we have used $K_{WWW}=2.45$,
\textit{i.e}, the value obtained for inclusive $WWW$ production, instead of the much smaller
value $K_{WWW}=1.29$ one obtains for $WWW$ production with a jet veto. It is very
possible that we may have over-estimated this background, but we choose
to err on the conservative side in our assessment of the discovery
prospects of the HL-LHC, the subject of the next section.

\section{Discovery prospects at the HL-LHC}
\label{sec:reach}

\begin{figure}[htbp]
\begin{center}
\includegraphics[width=15cm,clip]{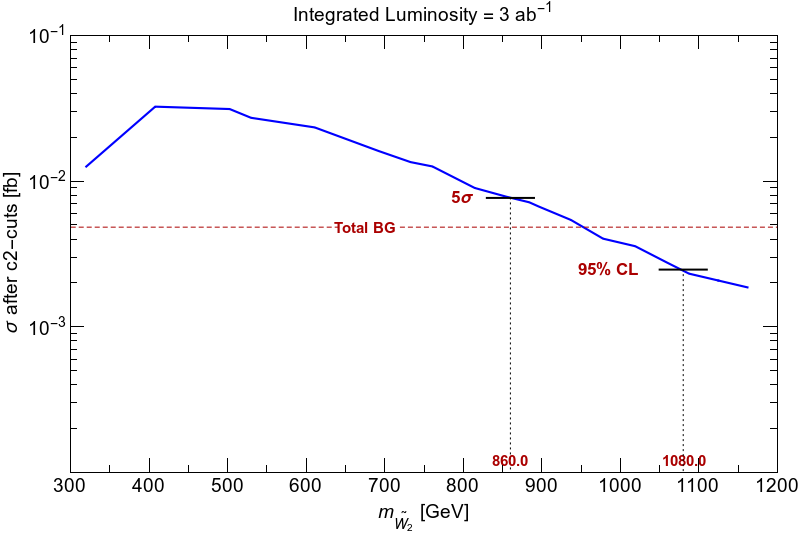}
\end{center}
\caption{Cross section for SSdB production 
after {\bf C2} cuts versus $m(wino)$ at the LHC with 
$\sqrt{s}=14$ TeV.  We show the $5\sigma$ and 95\% CL reach 
assuming a HL-LHC integrated luminosity of 3 ab$^{-1}$.
\label{fig:reach}}
\end{figure}

In Fig. \ref{fig:reach}, we show the total same sign dilepton signal
rate after our final analysis cuts, {\bf C2}, as a function of the wino
mass, $m_{\tw_2}$, (solid blue curve) along with the total SM background
(denoted by the dotted red line).  We also compute the reach for
$5\sigma$ discovery and 95\% CL exclusion for the HL-LHC (using Poisson
statistics) with a data sample of 3 ab$^{-1}$.  We find that the
$5\sigma$ discovery reach extends to $m_{\tw_2}\sim 860$ GeV, while the
95\% CL exclusion reach extends to $m_{\tw_2}\sim 1080$ GeV.  As
stressed previously, although the model line we have used includes the
assumption of gaugino mass unification, our projected reach does not
depend on this assumption, but only on $M_2\gg |\mu|$, as expected in
natural SUSY.  In models with gaugino mass unification, the $5\sigma$
(95\% CL) reach in $m_{\tw_2}$ correspond to a reach (exclusion) in
terms of the unified gaugino mass $m_{1/2}$ of $\sim 1010$ (1280) GeV.
In terms of the comparable reach in terms of $m_{\tg}$, these correspond
to $m_{\tg}\sim 2430$ (3000) GeV. These values may be compared to the
$5\sigma$ 3 ab$^{-1}$ HL-LHC for direct gluino pair production of
$m_{\tg}\sim 2800$ GeV obtained in Ref.~\cite{mgluino}.
Although we do not show it on the figure, we mention that with the hard
{\bf C2} cuts, the discovery reach of the LHC extends to 500~GeV
(720~GeV) for an integrated luminosity of 300~fb$^{-1}$ (1~ab$^{-1}$),
while the corresponding 95\%CL exclusion extends to 780~GeV
(980~GeV). It is worth keeping in mind that especially for the
300~fb$^{-1}$ case, somewhat softer analysis cuts~\cite{lhc,lhcltr} may
be better suited for optimizing the LHC reach.

\begin{figure}[htbp]
\begin{center}
\includegraphics[width=15cm,clip]{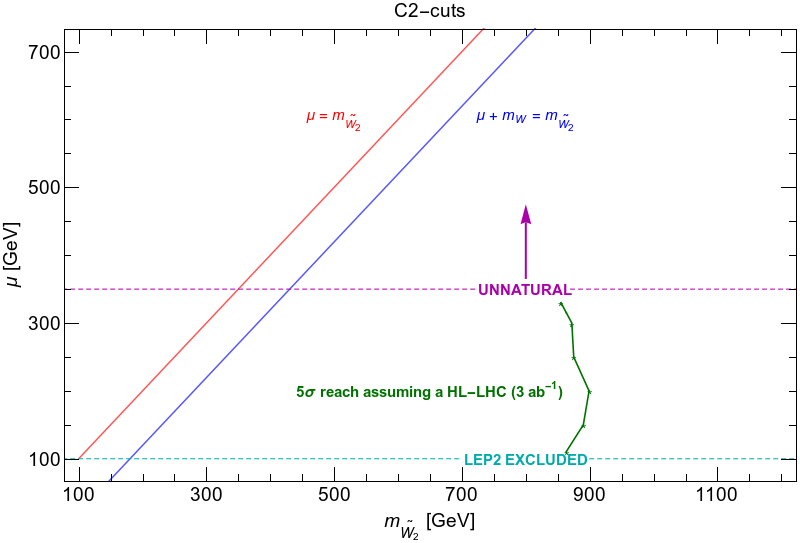}
\end{center}
\caption{Discovery reach in the SSdB channel at the HL-LHC in the
  $m_{\tw_2}$ vs. $\mu$ plane.
\label{fig:simp_model}}
\end{figure}

The key mass relation for the SSdB signature is that $|\mu |\ll M_2$.
It is therefore interesting to explore our discovery reach beyond our
benchmark assumption of $|\mu| = 150$ GeV.  In
Fig.~\ref{fig:simp_model}, we denote the (3 ab$^{-1}$) HL-LHC
($5\sigma$) discovery reach in the $\mu$-$M_2$ plane by the green solid
line in the vicinity of $m_{\tw_2} \simeq 850-900$~GeV.  As expected the
reach is only weakly sensitive to the higgsino mass.
The red diagonal line in Fig.~\ref{fig:simp_model} shows where $\mu
=m_{\tw_2}$. Above this line the SSdB signature arises from higgsino pair
production and subsequent decays to winos; but it would have a much smaller
rate because~(1)~the higgsino cross section is smaller than the wino
cross section, and (2)~dilution of the signal from higgsino decays to
binos (if these are accessible). Below the blue diagonal line in
Fig.~\ref{fig:simp_model} denotes the region where $\tw_2\to
\tz_{1,2}+W$ or $\tz_4\to \tw_1+W$ decays can occur, leading the the
SSdB final state, with on-shell $W$s. Close to this line and for 
not-too-large $m_{\tw_2}$, though, the
same sign dilepton events would not necessarily be clean as the large
wino-higgsino mixings would lead to sizeable mass gaps and concomitant
harder debris from the decay of the lighter inos. As $\mu$ increases,
the model becomes increasingly unnatural, with a value $\mu >350$
(indicated by a magenta dashed line) corresponding to electroweak
fine-tuning measure $\Delta_{EW}>30$.  The natural SUSY region is the
region below this horizontal line.

\section{SSdB SUSY event characteristics} \label{sec:char}

We have already illustrated the $\eslt$ and lepton transverse momentum
distributions after all cuts in Fig.~\ref{fig:met} and
Fig.~\ref{fig:ptl1}, respectively.  We saw that while the $\eslt$ 
distribution from signal
emerges from the background for $\esltTwo > 250$~GeV, this distribution is
typically backed up against the cut. Although the distribution may
harden somewhat with increasing wino mass, we saw that the observability
of the signal becomes rate limited by the time we reach $m_{\tw_2} =
860$~GeV, so wino events would typically have $\esltTwo \sim
250-500$~GeV. The lepton $p_T$ distributions peak at 200-250~GeV for the
hard lepton and 50-100~GeV for the second lepton, independent of the
wino mass. This should not be very surprising because the leptons are
produced at the end of a cascade decay chain, so the $p_{T\ell}$
distributions are only altered by the changes in the boost of the
daughter $W$ bosons which share the parent wino energy with the (nearly
invisible) higgsinos.

\begin{figure}[tbp]
\begin{center}
\includegraphics[width=7.5cm,clip]{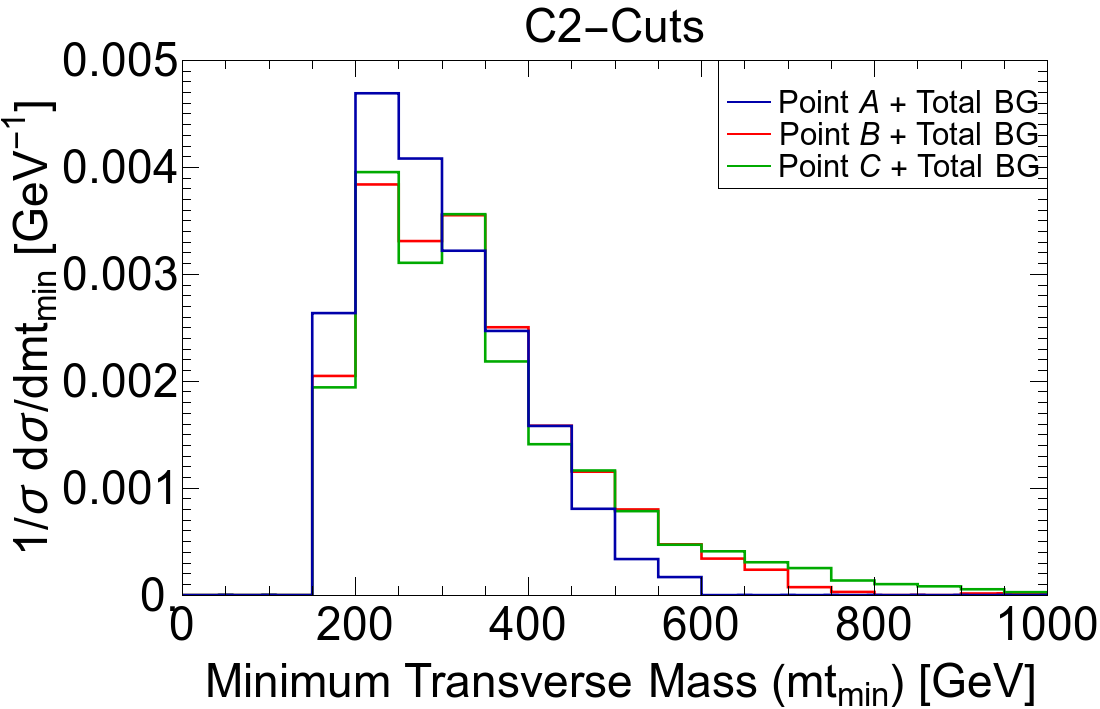} 
  \includegraphics[width=7.5cm,clip]{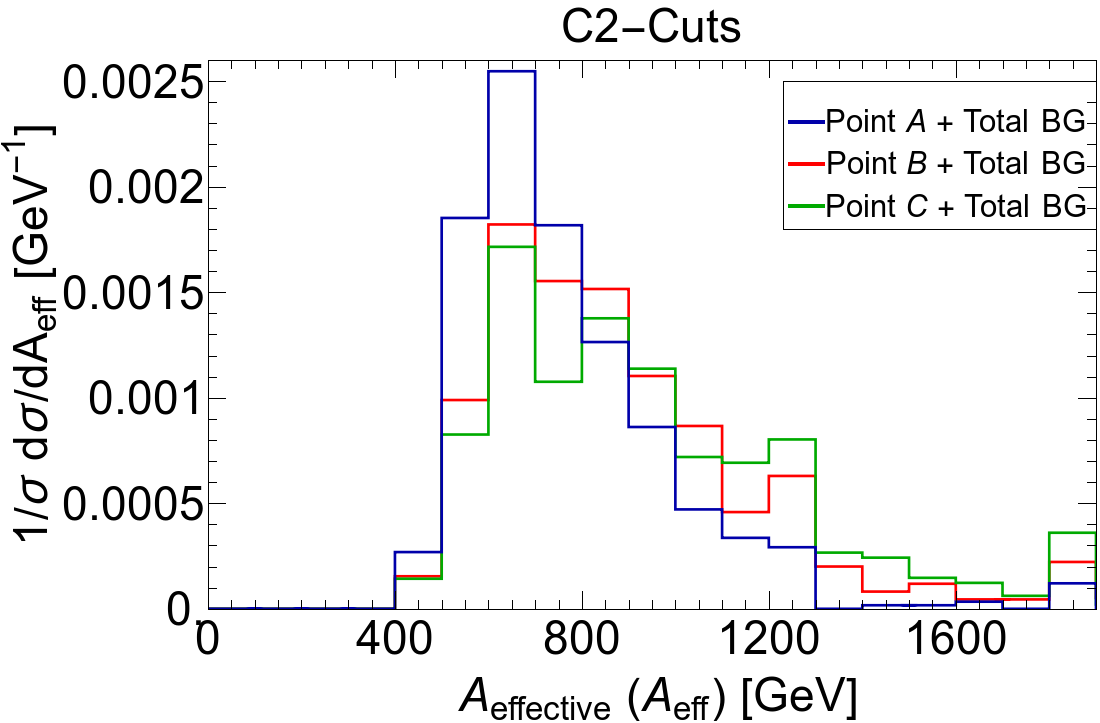} 
\includegraphics[width=7.5cm,clip]{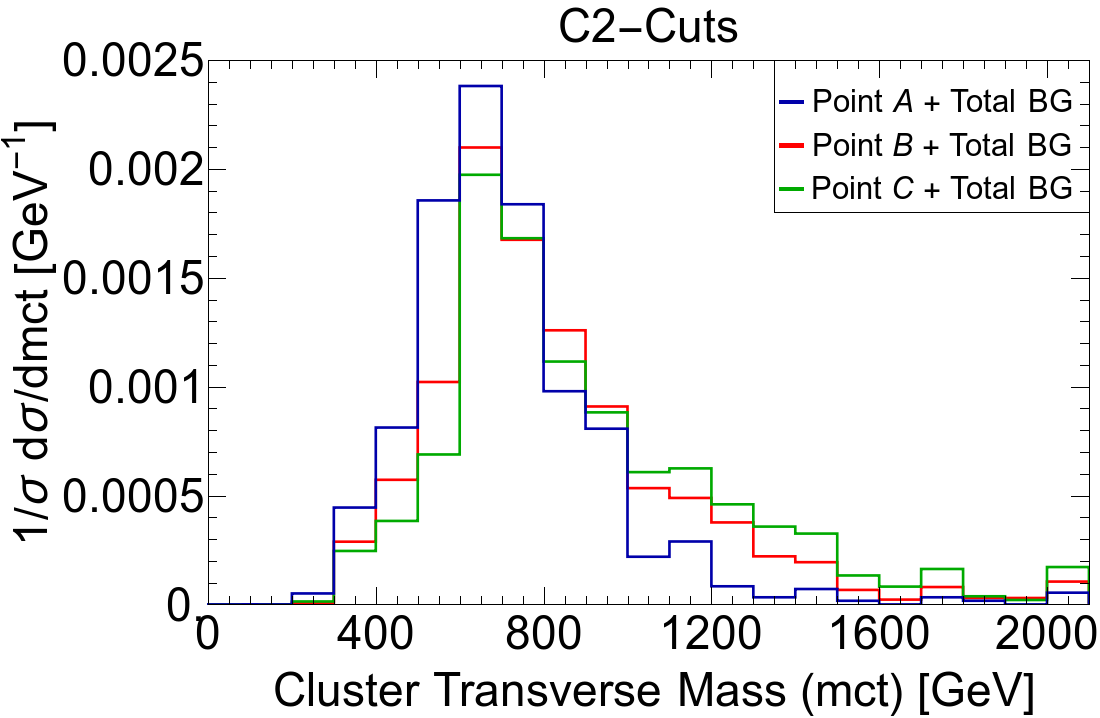}
\includegraphics[width=7.5cm,clip]{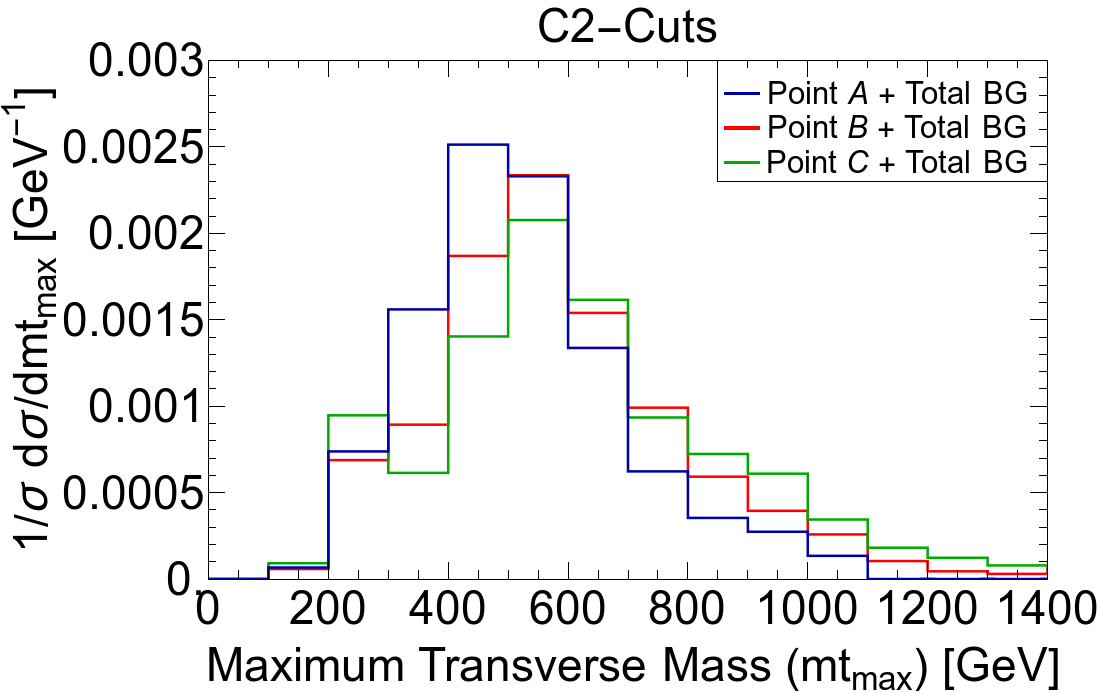}
\end{center}
\caption{Distributions of $m_T^{min}$ (top left), $A_{eff}$ (top right),
$m_{CT}$ (bottom left) and $m_T^{max}$ (bottom right) from the SUSY SSdB
signal plus SM backgrounds after {\bf C2} cuts for the three benchmark
cases {\tt Point A}, {\tt Point B}, and {\tt Point C} introduced earlier
in the text. We have normalized these distributions to all have the same
area.
\label{fig:dist}}
\end{figure}

To further characterize the nature of the SSdB events from SUSY, and
to see if we can gain some sensitivity to the wino mass from the
kinematic properties of these events, we have examined several
kinematic variables: $A_{eff}$, $m_T^{min}$ (which entered the {\bf C1}
cuts), its sibling $m_T^{max}$, $m_{T2}$, $m_{CT}$ and $m_{\ell\ell}$, where
\begin{equation}
\label{eq:A_eff}
A_{eff} = \eslt + \sum_i^{n(j)} p_T(j_i) + p_T(\ell_1) + p_T(\ell_2),
\nonumber
\end{equation}
and $m_{CT}$ is the cluster transverse mass given by
\begin{equation}
  m_{CT}^2= m_{CT}^2= \left(\eslt+\sqrt{\vec{p}_{T\ell\ell}^{\,2} + m_{\ell\ell}^2}   \right)^2 -
     (\!\!\vec{\,\,\eslt}+\! \vec{\,p}_{T\ell\ell})^2. \nonumber
  \end{equation}
  In Fig.~\ref{fig:dist},
we show the normalized distributions of $m_T^{min}$ (because it enters
our analysis cuts) together with those of $A_{eff}$, $m_{CT}$, and
$m_T^{max}$, the larger of the transverse masses of the lepton and
$\eslt$.  These are the distributions whose
{\em shapes} show the most sensitivity to the wino mass for the
three benchmark SUSY cases introduced above.
We see that even for these three cases with a fairly wide separation of
wino masses, the shapes of the distributions are qualitatively quite
similar, with perhaps the $m_T^{max}$ distribution showing the greatest
sensitivity to the parent wino mass. As we noted in the discussion of
Fig.~\ref{fig:ptl1}, the wino mass has a relatively small effect on the
kinematics of signal events, affecting only the boost of the $W$ bosons.
While these (quite correlated) distributions show some differences,
especially in the tails of the distributions which correspond to relatively
low numbers of signal events, we will see below that because the signal
rate can be predicted with good precision, the event rate for the SSdB
signal offers a much better handle on the wino mass. We stress, though,
that the kinematic properties of these events are nonetheless useful for
validating the signal origin, and could potentially serve as ingredients
in an artificial neural network stew.

\begin{figure}[tbp]
\begin{center}
\includegraphics[width=15cm,clip]{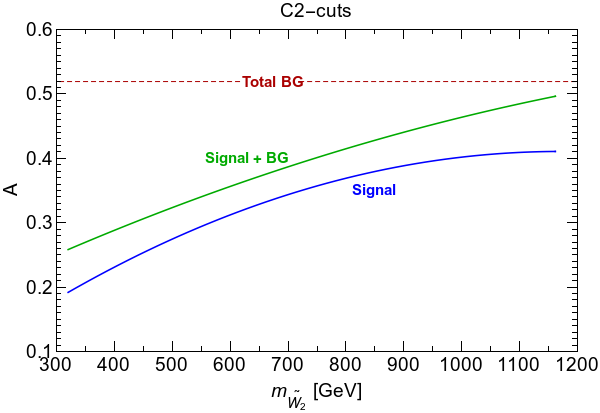}
\end{center}
\caption{Same-sign dilepton charge asymmetry 
from signal-plus-background vs. 
$m_{\tw_2}$ from SUSY same-sign diboson production 
after {\bf C2} cuts versus $m_{\tw_2}$ at LHC with $\sqrt{s}=14$ TeV.
The statistical error with which the charge asymmetry can be determined
is $\sim \pm 0.1$ is $m_{\tw_2}\alt 800$~GeV.
\label{fig:A}}
\end{figure}

The charge asymmetry $$A=\frac{n(++)-n(--)}{n(++)+n(--)}$$ of clean
same sign dilepton events (which, of course, includes both signal and
background events) provides yet another handle for validating the wino origin
of any signal.  We show a fit to the expected $A$ values (our simulated
sample had considerable statistical fluctuations) for
signal-plus-background events versus $m_{\tw_2}$ in Fig. \ref{fig:A},
together with the expected background value. The charge asymmetry arises
because there are more up-type than down-type valence quarks in a
proton. The importance of valence quark collisions for wino pair
production processes increases with wino mass, so we expect the
asymmetry to also increase with $m_{\tw_2}$. This is indeed borne out in
the figure where we see that the expected asymmetry ranges from 0.2 for
$m_{\tw_2}$ as low as $\sim 300$ GeV to 0.4 for $m_{\tw_2}\sim 1000$
GeV.\footnote{The asymmetry of the background is even larger because the
$W^\pm W^\pm jj$ component of the background, though subdominant, has
contributions from collisions of {\em two} valence quarks.}
Unfortunately, the measured charge asymmetry does not provide as good of a
wino mass determination as one might naively suppose from looking at the
figure. The reason is that because of the relatively low total event
rate, even with 3~ab$^{-1}$, the statistical error on its measurement is
$\sim \pm 0.1$ for $m_{\tw_2} < 800$~GeV, which corresponds to a wino
mass uncertainty of $\sim 300$ GeV. We nevertheless stress that a
determination of the charge asymmetry provides a consistency check of
wino origin of the SSdB signal if $m_{\tw_2}$ can be extracted from the
total event rate.  An examination of this extraction is the subject of the next
section.

\section{Measurement of the wino mass in the SSdB channel}
\label{sec:mass}

We saw that while experiments at the HL-LHC would be able to discover
winos with masses up to 860~GeV and to exclude these out to 1100~GeV
if no excess is seen, the determination of its mass from the kinematic
properties of the signal event proved rather difficult. We traced this to
the fact that the leptons were produced only at the end of a cascade so that
the sensitivity to the mass of the parent winos is correspondingly reduced.

In principle, it should also be possible to determine the wino mass from
the {\em rate} with which the signal events are produced. This is
particularly true in this case because the cross section for wino
production can be rather precisely computed for the case of natural SUSY
(for which the heavier inos are expected to be nearly pure gauginos) and
depends on just the wino mass. We also saw in
Sec.~\ref{ssec:signal-branching} that, at least for $m_{\tw_2}>
500$~GeV, the natural SUSY branching fraction for wino decays to $W$ is
 $0.25\pm 0.02$ with conservative error bars.\footnote{As
we have already noted, the observation of a signal in the clean, same
sign dilepton channel already points to light higgsinos and much heavier EW
gauginos. Additional circumstantial evidence for light higgsinos could,
for instance, come from the observation of monojet plus soft dilepton
events, which must be present at observable rates if
$m_{\tz_2}-m_{\tz_1} \agt 10$~GeV and higgsinos are not much heavier
than 220-240~GeV.}  The determination of the SSdB signal rate after {\bf
C2} cuts shown in Fig.~\ref{fig:reach} thus provides a plausible mass
measurement strategy, because, to a good approximation, the observed
number of events depends only on the wino mass.

For example, for our assumed benchmark point, {\tt Point B}, and using
{\bf C2} cuts,  with 3 ab$^{-1}$ we expect a total of  $63 \pm 8$ events
(see Table~\ref{tab:bg}), where the error bar is purely statistical.
Since we would estimate the signal cross section by taking the
observed number of events and subtracting the expected number of
background events, this $\pm$ 8 events corresponds to a $\approx$ 16\%
measurement of the cross section, which, as one can find by examining
the cross section after {\bf C2} cuts (as in Fig.~\ref{fig:reach})
corresponds to a measurement of $m_{\tw_2}\sim 690\pm 35$ GeV,  
which represents a better than 5\% measurement of the wino mass.

This precision is possible when we consider statistical
errors alone.  There is also a systematic error arising from the theory
uncertainty on the cross section, uncertainties on the wino decay
branching ratios, uncertainties on the efficiencies for events passing
cuts, uncertainties on the reconstruction efficiencies, etc.  Since the
current uncertainty ($\sim 10$\% in the production cross section) mostly
arises from the uncertainties in the parton distributions which will
undoubtedly be well-measured by the time this analysis is done, and the
lepton detection efficiencies will also be well understood, we expect the main
systematic will arise from the squared wino branching fraction, which as
we have already noted is $\alt 16$\%. Conservatively taking the total
systematic to be $\sim 20$\%, then our error on the wino mass for {\tt
Point B} increases to $\approx$ 50 GeV. Even if the total systematic
error on the cross section is 30\%, then the combined statistical and
systematic error on the mass is $\approx$ 70 GeV, which is about a 10\%
measurement of the wino mass.  If our background is underestimated by a
factor of two, our measurement of the wino mass will be biased by
$\approx$ 70 GeV toward lower values; if it is over-estimated by a
factor of two, then our measurement will be biased by $\approx$ 35 GeV
toward higher values.

We can still make a good mass measurement for large values of the wino
mass; for instance, the purely statistical error on the mass
measurement is still only $\approx$ 10\% for a 1 TeV wino (although
there is no $5\sigma$ signal).  However for these larger mass values
with their correspondingly smaller signal cross sections, very precise
determinations of the background cross section become increasingly
important. Presumably, these will be experimentally determined by an
extrapolation into the signal region by the time the HL-LHC
accumulates 3~ab$^{-1}$ of data. Our point is that better than 10\%
determination of the wino mass will be possible if the SSdB signal
from natural SUSY is detected at the HL-LHC

\section{Conclusions} 
\label{sec:conclude}

In this paper we have re-visited and explored aspects of the SSdB
signature, which is a powerful channel for discovering natural SUSY
models with $|\mu| \ll M_2$, especially if $M_3$ is larger than in
unified models.  This signature arises from wino pair production, $pp
\to\tw_2\tz_4$, followed by wino decays to $W$ bosons plus quasi-visible
higgsinos.  Thus, the signal consists of
$\ell^\pm\ell^{\prime\pm}+\eslt$ events which are distinct from
same-sign dilepton events from gluino/squark production in that they are
relatively free of hard jet activity. We emphasize that the SSdB search
channel offers a probe of natural SUSY -- indeed of all SUSY models with
light higgsinos -- that is independent of any signals from gluino pair
or top-squark pair production.  The SSdB channel is especially useful
because (i) SM backgrounds for such a signature are tiny and (ii) this
type of signature is not expected in many previously studied
``unnatural'' SUSY models, such as mSUGRA/CMSSM, where the opposite mass
hierarchy, $M_2<|\mu |$, and $M_1< M_2$ is expected.

We have evaluated several new background contributions to the SSdB
signature including $WWjj$ production, $4t$ production, and $3W$
production.  We find these new background reactions can be suppressed
beyond the previously examined {\bf C1} cuts by an additional jet veto
$n(jets)\le 1$ and a hardened $\esltTwo$ cut at a modest cost to the
signal.  The surviving signal rate should be observable at HL-LHC with
3 ab$^{-1}$ of integrated luminosity over a large range of wino mases.
After our {\bf C2} analysis cuts, the HL-LHC $5\sigma$ reach (95\%CL
exclusion) extends out to $m_{\tw_2}=860$ GeV (1080 GeV).  
We show that a
determination of the clean same sign dilepton event rate allows a
better than 10\% measurement of the wino mass over the entire range of
masses for which experiments at the HL-LHC will be able to discover a
wino in this channel. A measurement of the like-sign dilepton lepton
charge asymmetry will test the consistency of the wino origin of the
signal. If gluinos are also discovered at the HL-LHC, experiments will
be able to probe whether or not gaugino masses arise from a common
mass at $Q \simeq M_{\rm GUT}$ at the 10\% level~\cite{mgluino}.  We
encourage continued experimental scrutiny of the clean same sign dilepton +
$\eslt$ channel as the integrated luminosity at the LHC goes beyond
$\sim 100$~fb$^{-1}$. 

\section*{Acknowledgments}

This work was supported in part by the US Department of Energy, 
Office of High Energy Physics and was performed in part at the 
Aspen Center for Physics, which is supported by 
National Science Foundation grant PHY-1607611.

%

%

\begin{thebibliography}{99}
\small
\bibitem{lhc_mgl} The ATLAS collaboration [ATLAS Collaboration],
  ATLAS-CONF-2017-022;
T.~Sakuma [CMS Collaboration],
  PoS LHCP {\bf 2016} (2017) 145
  [arXiv:1609.07445 [hep-ex]].
%
\bibitem{lhc_mt1} The ATLAS collaboration [ATLAS Collaboration],
  ATLAS-CONF-2017-037;
A.~M.~Sirunyan {\it et al.} [CMS Collaboration],
  arXiv:1706.04402 [hep-ex].
%
\bibitem{craig} N.~Craig,
  arXiv:1309.0528 [hep-ph].
%
%
\bibitem{bg} R.~Barbieri and G.~F.~Giudice, 
Nucl.\ Phys.\ B {\bf 306}, 63 (1988).
%
\bibitem{papp} R.~Kitano and Y.~Nomura, Phys. Rev. D {\bf 73} (2006)
  095004; M.~Papucci, J.~T.~Ruderman and A.~Weiler, JHEP {\bf 1209}
  (2012) 035.
%
\bibitem{reduces} H.~Baer, V.~Barger and D.~Mickelson, 
Phys.\ Rev.\ D {\bf 88}, (2013) 095013.
%
\bibitem{am} A.~Mustafayev and X.~Tata, Indian J. Phys. {\bf 88} (2014)
991.
%
\bibitem{seige} H.~Baer, V.~Barger, D.~Mickelson and M.~Padeffke-Kirkland,
  Phys.\ Rev.\ D {\bf 89} (2014) 115019.
%
\bibitem{rns} H.~Baer, V.~Barger, P.~Huang, D.~Mickelson, A.~Mustafayev and X.~Tata,
  Phys.\ Rev.\ D {\bf 87} (2013) 11,  115028.
%
\bibitem{ltr} H.~Baer, V.~Barger, P.~Huang, A.~Mustafayev and X.~Tata,
  Phys.\ Rev.\ Lett.\  {\bf 109} (2012) 161802.
%
\bibitem{helhc} For a recent overview of upper bounds on stop and gluino
masses in a variety of models from the  requirement $\Delta_{EW}< 30$, see
Sec. 5 of H.~Baer, V.~Barger, J.~Gainer, H.~Serce and X.~Tata,
arXiv:1708.09054 [hep-ph].

%
\bibitem{sundrum} C.~Brust, A.~Katz, S.~Lawrence and R.~Sundrum,
  JHEP {\bf 1203} (2012) 103.

%
\bibitem{ross} L.~Girardello and M.~T.~Grisaru, Nucl. Phys. B {\bf 194}
  (1982) 75, and recently re-emphhasized by G.~G.~Ross,
  K.~Schmidt-Hoberg and F.~Staub, Phys. Lett. B {\bf 759} (2016) 110.
%
\bibitem{other} T.~Cohen, J.~Kearney and M.~Luty, Phys. Rev. D {\bf 91}
  (2014) 075004; A.~Nelson and T.~Roy, Phys. Rev. Lett. {\bf 114} (2015)
  201802; S.~P.~Martin, Phys. Rev. D {\bf 92} (2015) 035004.


\bibitem{eenz} J.~R.~Ellis, K.~Enqvist, D.~V.~Nanopoulos and
  F.~Zwirner, 
  Phys.\ Lett.\ A {\bf 1}, 57 (1986).
%
  %
\bibitem{Baer:2015tva} 
  H.~Baer, V.~Barger, P.~Huang, D.~Mickelson, M.~Padeffke-Kirkland and X.~Tata,
  Phys.\ Rev.\ D {\bf 91}, no. 7, 075005 (2015).
%
\bibitem{winosearch} CMS Collaboration, CMS-PAS-SUS-17-004; ATLAS
Collaboration, ATLAS-CONF-2017-039
%
\bibitem{pqww} R.~D.~Peccei and H.~R.~Quinn,
  Phys.\ Rev.\ Lett.\  {\bf 38} (1977) 1440;
R.~D.~Peccei and H.~R.~Quinn,
  Phys.\ Rev.\ D {\bf 16} (1977) 1791;
S.~Weinberg,
  Phys.\ Rev.\ Lett.\  {\bf 40} (1978) 223;
F.~Wilczek,
  Phys.\ Rev.\ Lett.\  {\bf 40} (1978) 279.
%
\bibitem{bbc} K.~J.~Bae, H.~Baer and E.~J.~Chun,
  Phys.\ Rev.\ D {\bf 89} (2014) no.3,  031701;
K.~J.~Bae, H.~Baer and E.~J.~Chun,
  JCAP {\bf 1312} (2013) 028.



%
\bibitem{mono} H.~Baer, A.~Mustafayev and X.~Tata, 
  mono-photons from light higgsino pair production at LHC14,'' Phys.\
  Rev.\ D {\bf 89} (2014) no.5, 055007. See also, C.~Han, A.~Kobakhidze,
  N.~Liu, A.~Saavedra, L.~Wu and J.~M.~Yang, JHEP {\bf 1402} (2014) 049;
  P.~Schwaller and J.~Zurita, 
  the LHC,'' JHEP {\bf 1403}, 060 (2014).
 %
\bibitem{llj} 
Z.~Han, G.~D.~Kribs, A.~Martin and A.~Menon,
  Phys.\ Rev.\ D {\bf 89} (2014) no.7,  075007;
H.~Baer, A.~Mustafayev and X.~Tata,
  Phys.\ Rev.\ D {\bf 90} (2014) no.11,  115007.



%
\bibitem{bmt} H.~Baer, V.~Barger, D.~Mickelson, A.~Mustafayev and X.~Tata,
  JHEP {\bf 1406} (2014) 172.
%
\bibitem{ilcgroup} H.~Baer, M.~Berggren, K.~Fujii, S.~L.~Lehtinen, J.~List, T.~Tanabe and J.~Yan,
  PoS ICHEP {\bf 2016} (2016) 156
  [arXiv:1611.02846 [hep-ph]].


%
\bibitem{mgluino} H.~Baer, V.~Barger, J.~S.~Gainer, P.~Huang, M.~Savoy, D.~Sengupta and X.~Tata,
\epjc {77} (2017) 499  [arXiv:1612.00795 [hep-ph]].
%
\bibitem{stop} H.~Baer, V.~Barger, N.~Nagata and M.~Savoy,
  Phys.\ Rev.\ D {\bf 95} (2017) no.5,  055012.
  %
\bibitem{lhc} H.~Baer, V.~Barger, P.~Huang, D.~Mickelson, A.~Mustafayev, W.~Sreethawong and X.~Tata,
  JHEP {\bf 1312} (2013) 013
   Erratum: [JHEP {\bf 1506} (2015) 053].
   %


%
\bibitem{baris} B.~Altunkaynak, H.~Baer, V.~Barger and P.~Huang,
  Phys.\ Rev.\ D {\bf 92} (2015) no.3,  035015.
%
\bibitem{cms_llj} CMS Collaboration, see talk by
J. Hirschauer, Pheno2017 meeting, Pittsburgh, PA, May, 2017.

\bibitem{lhcltr} H.~Baer, V.~Barger, P.~Huang, D.~Mickelson, A.~Mustafayev, W.~Sreethawong and X.~Tata,
  Phys.\ Rev.\ Lett.\  {\bf 110} (2013) no.15,  151801.
  %

%
\bibitem{ssdl} R.~M.~Barnett, J.~F.~Gunion and H.~E.~Haber,
  UCD-88-30, LBL-26204, SCIPP-88-36; 
H.~Baer, X.~Tata and J.~Woodside,
  Phys.\ Rev.\ D {\bf 41} (1990) 906;
H.~Baer, X.~Tata and J.~Woodside,
  Phys.\ Rev.\ D {\bf 45} (1992) 142;
R.~M.~Barnett, J.~F.~Gunion and H.~E.~Haber,
  Phys.\ Lett.\ B {\bf 315} (1993) 349.
%
\bibitem{nuhm2} D.~Matalliotakis and H.~P.~Nilles,
  Nucl.\ Phys.\ B {\bf 435} (1995) 115;
M.~Olechowski and S.~Pokorski,
  Phys.\ Lett.\ B {\bf 344} (1995) 201;
P.~Nath and R.~L.~Arnowitt,
  Phys.\ Rev.\ D {\bf 56} (1997) 2820;
J. Ellis, K. Olive and Y. Santoso, Phys. Lett. {\bf B539} (2002) 107;
J. Ellis, T. Falk, K. Olive and Y. Santoso, 
Nucl. Phys. {\bf B652} (2003) 259;
H.~Baer, A.~Mustafayev, S.~Profumo, A.~Belyaev and X. Tata, 
JHEP{\bf 0507} (2005) 065.
%
\bibitem{btag} U.~Chattopadhyay, A.~Datta,  A.~Datta,  A.~Datta and
D.P.~Roy, Phys. Lett. B {\bf 493} (2000) 127; P.~Mercadante, J.~K.~Mizukoshi
and X.~Tata, Phys. Rev. D {\bf 72} (2005) 035009; R.H.K.~Kadala,
P.~Mercadante, J.~K.~Mizukoshi and X.~Tata, \epjc
{\bf 56} (2008) 511.
%
\bibitem{multi} H.~Baer, V.~Barger, M.~Savoy and X.~Tata,
  Phys.\ Rev.\ D {\bf 94} (2016) no.3,  035025.
%
\bibitem{ngmm} H.~Baer, V.~Barger, H.~Serce and X.~Tata,
  Phys.\ Rev.\ D {\bf 94} (2016) no.11,  115017.
%
\bibitem{mini} H.~Baer, V.~Barger, M.~Savoy, H.~Serce and X.~Tata,
  arXiv:1705.01578 [hep-ph].
%
\bibitem{madgraph} J.~Alwall, M.~Herquet, F.~Maltoni, O.~Mattelaer and
  T.~Stelzer,
  JHEP {\bf 1106} (2011)
  128; J.~Alwall, R.~Frederix, S.~Frixone, V~Herschi, F.~Maltoni,
  O.~Mattelaer, H.-S.~Shao, T.~Stelzer, P.~Torrielli and M.~Zaro, JHEP
  {\bf 1407} (2014) 079.
%
\bibitem{pythia} T.~Sj\"{o}strand {\it et al.},
  Comput.\ Phys.\ Commun.\  {\bf 191} (2015) 159.
%
\bibitem{delphes} J.~de Favereau {\it et al.} [DELPHES 3 Collaboration],
  JHEP {\bf 1402} (2014) 057.
  %
\bibitem{prospino} W.~Beenakker, R.~Hopker and M.~Spira,
  hep-ph/9611232.
%
\bibitem{isajet} F.~E.~Paige, S.~D.~Protopopescu, H.~Baer and X.~Tata,
  hep-ph/0312045.
%
\bibitem{wss} H.~Baer and X.~Tata, {\em Weak Scale Supersymmetry},
(Cambridge University Press, 2006)

%
  %
\bibitem{Czakon:2013goa} 
  M.~Czakon, P.~Fiedler and A.~Mitov,
  Phys.\ Rev.\ Lett.\  {\bf 110}, 252004 (2013).
 %
  \bibitem{Bevilacqua:2012em} 
  G.~Bevilacqua and M.~Worek,
  JHEP {\bf 1207}, 111 (2012).
%
\bibitem{Campbell:2011bn} 
  J.~M.~Campbell, R.~K.~Ellis and C.~Williams,
  JHEP {\bf 1107}, 018 (2011).
  %
  \bibitem{Campbell:2012dh} 
  J.~M.~Campbell and R.~K.~Ellis,
  JHEP {\bf 1207}, 052 (2012).
  %
  \bibitem{Kardos:2011na} 
  A.~Kardos, Z.~Trocsanyi and C.~Papadopoulos,
  Phys.\ Rev.\ D {\bf 85}, 054015 (2012).
  %
  \bibitem{Melia:2010bm} 
  T.~Melia, K.~Melnikov, R.~Rontsch and G.~Zanderighi,
  JHEP {\bf 1012}, 053 (2010).
  %
  \bibitem{Yong-Bai:2016sal} 
  Y.~B.~Shen, R.~Y.~Zhang, W.~G.~Ma, X.~Z.~Li and L.~Guo,
  Phys.\ Rev.\ D {\bf 95}, no. 7, 073005 (2017).
%
  \bibitem{Cacciari:2008gp} 
  M.~Cacciari, G.~P.~Salam and G.~Soyez,
  JHEP {\bf 0804}, 063 (2008).
  %
  \bibitem{Cacciari:2011ma} 
  M.~Cacciari, G.~P.~Salam and G.~Soyez,
  Eur.\ Phys.\ J.\ C {\bf 72}, 1896 (2012).
  %
  \bibitem{CMS:2016kkf} 
  CMS Collaboration [CMS Collaboration],
  CMS-PAS-BTV-15-001.
  %
\bibitem{Cacciari:2007fd} 
  M.~Cacciari and G.~P.~Salam,
  Phys.\ Lett.\ B {\bf 659}, 119 (2008).
  %
\bibitem{ATLASb} S.~Corr\'ead, V. Kostioukine, J.~Lev\^eque,
  A.~Rozanov, J.~B.~de Vivie, ATLAS Note, ATLAS-PHYS-2004-006, and 
V. Kostioukine, ATLAS Note, ATLAS-PHYS-2003-033.
  %
\bibitem{mct} 
  V.~D.~Barger, T.~Han and R.~J.~N.~Phillips,
  Phys.\ Rev.\ D {\bf 36}, 295 (1987).
%
\bibitem{mt2} 
  C.~G.~Lester and D.~J.~Summers,
  Phys.\ Lett.\ B {\bf 463}, 99 (1999); 
A.~Barr, C.~Lester and P.~Stephens,
  J.\ Phys.\ G {\bf 29}, 2343 (2003).
%


\end{thebibliography}
\end{document}